\documentclass[aoas,preprint]{imsart}

\RequirePackage[OT1]{fontenc}
\RequirePackage[utf8]{inputenc}
\RequirePackage{latexsym,amsmath,amsthm,amsfonts,graphicx,float}

\RequirePackage{natbib}
\RequirePackage[colorlinks,citecolor=blue,urlcolor=blue]{hyperref}
\RequirePackage{graphicx}
\DeclareGraphicsExtensions{.pdf,.jpg,.png,.jpeg}
\usepackage{hyperref}
\arxiv{arXiv:1505.05482}

\newcommand{\ve}[1]{{\mbox{\boldmath ${#1}$}}}  

\startlocaldefs
\numberwithin{equation}{section}
\theoremstyle{plain}

\endlocaldefs
\begin{document}

\begin{frontmatter}
\title{TPRM:  Tensor partition regression models with applications in   imaging biomarker detection}
\runtitle{Tensor partition regression models}

\begin{aug}
\author{\fnms{Michelle} \snm{F. Miranda}\thanksref{t1,m2,m1}\ead[label=e1]{michellemirandaest@gmail.com}},
\author{\fnms{Hongtu} \snm{Zhu}\thanksref{t3,m2,m3}\ead[label=e2]{hzhu5@mdanderson.org}}
\and \\ 
\author{\fnms{Joseph} \snm{G. Ibrahim}\thanksref{t2,m3}\ead[label=e3]{ibrahim@bios.unc.edu}},

\author{\fnms{for} \snm{the Alzheimer's Disease Neuroimaging Initiative}\thanksref{t4}} 

\thankstext{t1}{Dr. Miranda's research was partially supported by grant 2013/ 07699-0 and 2014/07254-0, Sao Paulo Research Foundation, and grant CA-178744.}
\thankstext{t2}{Dr. Ibrahim's research
was partially supported by  NIH grants \#GM 70335 and
P01CA142538.}
\thankstext{t3}{Dr.  Zhu  was partially supported by   NIH grant  MH086633,  NSF Grants  SES-1357666 and DMS-1407655, and a grant from Cancer Prevention Research Institute of Texas.}
\thankstext{t4}{Data used in preparation of this article were obtained from the Alzheimer’s Disease
Neuroimaging Initiative (ADNI) database (adni.loni.usc.edu). As such, the investigators
within the ADNI contributed to the design and implementation of ADNI and/or provided data
but did not participate in analysis or writing of this report. A complete listing of ADNI
investigators can be found at: http://adni.loni.usc.edu/wp-content/uploads/how\_to\_apply/ADNI\_Acknowledgement\_List.pdf.}
\runauthor{Miranda et al.}

\affiliation{ University of Texas MD Anderson Cancer Center\thanksmark{m2}, Universidade de S\~{a}o Paulo\thanksmark{m1}, and University of North Carolina at Chapel Hill\thanksmark{m3}}

\address{{\bf Michelle F. Miranda}\\
University of Texas MD Anderson Cancer Center and \\
Universidade de São Paulo, SP Brazil\\
\printead{e1}\\
{\bf Hongtu Zhu}\\
Department of Biostatistics\\
University of Texas MD Anderson Cancer Center and \\ 
University of North Carolina at Chapel Hill\\
\printead{e2}} 
\address{{\bf Joseph. G. Ibrahim}\\
Department of Biostatistics\\
University of North Carolina at Chapel Hill\\
\printead{e3}}
\end{aug}

\begin{abstract}
Medical imaging  studies have collected high dimensional imaging data to  identify imaging biomarkers 
for diagnosis, screening, and prognosis, among many others. These  imaging data  are often represented in the form of a multi-dimensional array, called a tensor.  The aim of this paper is to develop a tensor partition regression modeling (TPRM) framework  to establish a relationship between low-dimensional clinical outcomes (e.g., diagnosis)  and  high dimensional tensor covariates. 
Our TPRM is  a hierarchical model and   efficiently integrates  four components:  (i)  a partition model,  (ii)  a canonical polyadic decomposition model, (iii) a principal components model, and  (iv)  a generalized linear  model  with  a  sparse  inducing  normal  mixture  prior.  
This framework not only reduces ultra-high dimensionality  to a manageable level, resulting in efficient estimation, but also optimizes prediction accuracy in the search for informative sub-tensors.
Posterior computation proceeds via an efficient Markov chain Monte Carlo algorithm. Simulation shows that TPRM outperforms several other competing methods. We apply TPRM to predict disease status (Alzheimer versus control) by using structural magnetic resonance imaging data obtained from the Alzheimer's Disease Neuroimaging Initiative (ADNI) study. 
\end{abstract}


\end{frontmatter}

\section{Introduction}

Medical imaging studies have collected high dimensional imaging data (e.g., Computed Tomography (CT) and Magnetic Resonance Imaging (MRI)) to extract  
 information associated with the pathophysiology of various diseases. These information, or imaging biomarkers, could potentially aid detection and improve diagnosis, assessment of prognosis, prediction of response to treatment, and monitoring of disease status. Thus, efficient imaging biomarker extraction is crucial to the understanding of many disorders, including different types of cancer (e.g. lung cancer), and brain disorders such as Alzheimer's disease and autism, among many others.

 A critical challenge is  to  convert medical images into clinically useful information that can facilitate better clinical decision making \citep{gillies2015radiomics}.    
 Existing statistical methods are not always efficient for such conversion due to the  high-dimensionality of array images as well as their complex structure, such as spatial smoothness, correlation,  and heterogeneity. Although  a large family of regression methods has been developed for supervised learning of a scalar response (e.g. clinical outcome) \citep{Hastie2009,breiman84,Friedman1991,Zhang2010}, their computability and theoretical guarantee   are compromised by the ultra-high dimensionality of the imaging data covariates.  To address this challenge, many modeling strategies have been proposed to establish association
between high-dimensional array covariates and scalar response variables.

 The   first set of promising solutions is the  high-dimensional sparse regression (HSR) models, which often take high-dimensional imaging data as unstructured predictors. A key assumption of HSR is its sparse solutions.    HSRs not only suffer from  diverging spectra and noise accumulation in ultra-high dimensional feature space 
 \citep{FanFan2008a, Bickel2004}, but also their sparse solutions may  lack clinically meaningful information. 
 Moreover, standard HSRs  ignore  the inherent spatial structure of medical image, such as  spatial correlation and spatial smoothness.
 To address some limitations of HSRs,   a   family  of tensor regression models has been developed to preserve the tensor structure of imaging data, while achieving substantial dimension reduction   \citep{Zhou2013}.

The second set of solutions adopts   functional linear regression (FLR) approaches, which treat imaging data as functional predictors.  
However, since most existing FLR models  focus on one-dimensional  curves  \citep{MR2504202,Ramsay2005},
generalizations to two and higher dimensional images is far from trivial and requires substantial research \citep{Reiss2010}. 
   Most  estimation approaches of FLR  approximate the coefficient function as a linear combination of a set of fixed (or data-driven) basis functions.
For instance, most estimation methods  of FLR based on the fixed basis functions (e.g., tensor product wavelet) are required to solve an ultra-high dimensional optimization problem and 
can   suffer from  the same  limitations  as those of HSR.

 The third set of solutions  usually  integrates supervised (or unsupervised) dimension reduction techniques with various standard regression models.   Given the high dimension of imaging data,  it is imperative to use some dimension  reduction methods  to extract and select important `low-dimensional'  features, while eliminating most  noises  \citep{ Johnstone2009,Bair2006, FanFan2008a, Tibshirani2002,Krishnan2011}.
 Most of these methods first carry out an unsupervised dimension reduction step, often by principal component analysis (PCA), and then fit a regression model based on the top principal components \citep{Caffo2010}.  Recently,  for ultra-high tensor data, 
   unsupervised higher order tensor decompositions (e.g. parallel factor analysis and Tucker) have been extensively proposed to extract important information of neuroimaging data \citep{Martinez2004,Beckmann2005,Zhou2013}. These methods are intuitive and easy to implement, but features extracted from PCA and tensor decomposition can miss small and localized information that is relevant to the response. We propose a novel model that efficiently extracts these information, while performing dimension reduction and feature selection for better prediction accuracy.

The aim of this paper is to develop a novel modeling framework to extract imaging biomarkers from high-dimensional imaging data, denoted by ${\bf x}$,   to predict a  scalar response, denoted by $y$.  The scalar response  $y$ may include cognitive outcome, disease status, and the early onset of disease, among others. The imaging data provided by neuroimaging studies is often represented in the form of a multi-dimensional array, called a tensor.  
We  develop a novel Tensor Partition Regression Model (TPRM)  to establish an association between imaging tensor predictors and clinical outcomes.   
Our TPRM is  a hierarchical model with four  components, including  (i)  a partition model that divides the    high-dimensional tensor covariates   into  sub-tensor covariates;  (ii)  a canonical polyadic decomposition model that reduces the  sub-tensor covariates to    low-dimensional feature vectors;  (iii) a projection of these feature vectors into the space of the principal components, and  (iv) a generalized linear model with a sparse inducing normal mixture prior that is used to select informative feature vectors for predicting clinical outcomes. 
 Although the four components  of TPRM have been independently developed, the key novelty of TPRM lies in the integration of (i)-(iv) into a single framework for imaging prediction.
In particular, the first two components (i) and (ii) are designed to specifically address the three key features of neuroimaging data, including 
relatively low signal to noise ratio,  spatially clustered effect regions,  and the  tensor structure of imaging data.

In Section \ref{modelsec}, we introduce TPRM, the priors,  and a Bayesian estimation procedure. In Section \ref{simsec}, we use simulated data to compare the Bayesian decomposition with several competing methods. In Section \ref{rdsec}, we apply our model to the ADNI data set. This data set consists of 181 subjects with Alzheimer's disease and 221 controls and the correspondent covariates are MRI images of size $96 \times 96 \times 96$. In Section \ref{dissec}, we present some concluding remarks.

\section{Methodology}
\label{modelsec}

\subsection{Preliminaries}
\label{notation}

\begin{figure} [!htbp]
\begin{center}
\includegraphics[width=0.8\textwidth]{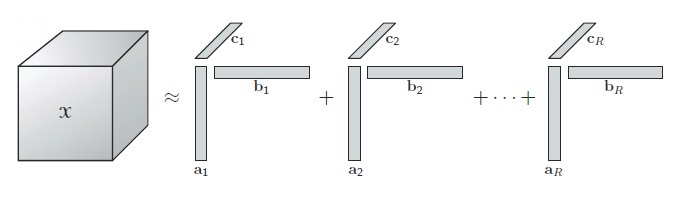}
\caption{ Figure copied from \citep{Kolda2009}. Panel (a) illustrates the CP decomposition of a three way array as a sum of R components of rank-one tensors, i.e.  $\mathcal{X} \approx \sum_{r=1}^R {\ve a}_r \circ {\ve b}_r \circ {\ve c}_r$. The approximation sign means that the right hand side is the the solution of $\min_{\tilde{\mathcal{X}}}\lVert \mathcal{X}-{\tilde{\mathcal{X}}}\Vert_2$, where $\Vert \cdot \Vert_2$ is the $L_2$ norm of tensors and  $\tilde{\mathcal{X}} = \sum_{r=1}^R {\ve a}_r \circ {\ve b}_r \circ {\ve c}_r$. This minimization problem can be written in a matricized version and solved using an alternating least squares (ALS) algorithm, please see \cite{Kolda2009} for details.}
\label{figureCP3D}
\end{center}
\end{figure}

We review a few basic facts about tensors \citep{Kolda2009}.   
A {\bf tensor} ${\bf x}=(x_{j_1 \ldots j_D})\in \mathbb{R}^{J_1\times \ldots \times J_D}$ is a multidimensional array, whose order $D$ is determined by its dimension. For instance, a vector is a tensor of order $1$ and a matrix is a tensor of order $2$. The {\bf inner product} between two tensors  $\mathcal{X}=(x_{j_1 \ldots j_D})$ and ${\mathcal{X}}'=({x}'_{j_1 \ldots j_D})$  in $\mathbb{R} ^{J_1\times \ldots \times J_D}$ is the sum of the product of their entries given  by 
$$
\langle \mathcal{X},  {\mathcal{X}}' \rangle=\sum_{j_1=1}^{J_1}\ldots\sum_{j_D=1}^{J_D}x_{j_1 \ldots j_D} {x}'_{j_1 \ldots j_D}. $$ 
 The {\bf outer product} between two vectors ${\ve a}^{(1)}=(a_{j_1}^{(1)}) \in \mathbb{R}^{J_1}$ and ${\ve a}^{(2)}=(a_{j_1}^{(2)}) \in \mathbb{R}^{J_2}$ is a matrix ${ M}=(m_{j_1j_2})$ of size $J_1 \times J_2$ with entries $m_{j_1j_2}=a_{j_1}^{(1)} a_{j_2}^{(2)}$. A tensor $\mathcal{X} \in \mathbb{R}^{J_1\times \ldots \times J_D}$ is a {\it rank one tensor} if it can be written as an outer product of $D$ vectors such that  $\mathcal{X}={\ve a}^{(1)} \circ {\ve a}^{(2)}\ldots \circ {\ve a}^{(D)}$, where ${\ve a}^{(k)}\in \mathbb{R}^{J_k}$ for $k=1, \ldots, D$. 
 Moreover, 
the canonical polyadic decompositic ({\bf CP decomposition}), also known as parallel factor analysis (PARAFAC),  factorizes a tensor into a sum of rank-one tensors such that 
\[
\mathcal{X} = \sum_{r=1}^R  \:{\ve a}_r^{(1)} \circ {\ve a}_r^{(2)} \circ \ldots \circ {\ve a}_r^{(D)},
\]
where  ${\ve a}_r^{(k)}=(a_{j_kr}^{(k)})\in \mathbb{R}^{J_k}$ for $k=1, \ldots, D$ and $r=1, \ldots, R$. See Figure \ref{figureCP3D} for an illustration of a 3D array.

It is convenient and assumed in this paper that the columns of the factor matrices are normalized to length one with weights absorbed into a diagonal matrix $\ve\Lambda=\mbox{diag}(\lambda_1, \ldots, \lambda_R)$ such that 
\begin{equation} 
\mathcal{X} = \sum_{r=1}^R \lambda_r \:{\ve a}_r^{(1)} \circ {\ve a}_r^{(2)} \circ \ldots \circ {\ve a}_r^{(D)} \equiv  \|\ve\Lambda; {\ve A}^{(1)}, \ldots, {\ve A}^{(D)}\|,  \label{TPRM1} \end{equation} 
where ${\ve A}^{(d)}=[{\ve a^{(d)}_1} {\ve a^{(d)}_2} \ldots {\ve a^{(d)}_R}]$ for $d=1,\ldots, D$. 

It is sometimes convenient to arrange the tensor $\mathcal{X}$ as a matrix. This arrangement can be done in various ways but we will rely on the following definition detailed in \cite{Kolda2009}. We define the {\bf mode-d matricized} version of $\mathcal{X}$ as
\[
{\ve X}_{(d)} = {\ve A}^{(d)}{\ve \Lambda} ({\ve A}^{(D)} \odot \ldots \odot {\ve A}^{(d+1)} \odot {\ve A}^{(d-1)} \odot \ldots \odot{\ve A}^{(1)})^T, 
\]
where $\odot$ denotes the Khatri–Rao product. Then, we can write the factor matrix corresponding to the dimension $d$ as a projection of ${\ve X}_{(d)}$ in the following way
\begin{equation}
{\ve A}^{(d)}={\ve X}_{(d)} ({\ve A}^{(D)} \odot \ldots \odot {\ve A}^{(d+1)} \odot {\ve A}^{(d-1)} \odot \ldots \odot{\ve A}^{(1)}) {\ve V^{\dagger}}{\ve \Lambda^{-1}},
\label{TPRM12}
\end{equation}
where 
${\ve V^{\dagger}}$ is the Moore-Penrose inverse of 
\[{\ve V}={\ve A}^{(1)T}{\ve A}^{(1)}*\ldots*{\ve A}^{(d-1)T}{\ve A}^{(d-1)}*{\ve A}^{(d+1)T}{\ve A}^{(d+1)}*\ldots*{\ve A}^{(D)T}{\ve A}^{(D)},\]
in which $* $ indicates the Hadamard product of matrices \citep{Kolda2009,kolda2006}.

We need the following notation throughout the paper.  
Suppose that we observe data $\{(y_i, \mathcal{X}_i, \ve z_i): i=1, \ldots, N\}$ from $N$ subjects, where the $\mathcal{X}_i$'s are tensor 
imaging data, $\ve z_i$ is a $p_z\times 1$ vector of scalar covariates, and $y_i$ is a scalar response, such as diagnostic status or 
clinical outcome. In the ADNI example, $N  = 402$ and $y_i$=1 if subject $i$ is a patient with Alzheimer's disease  and $y_i$=0 otherwise.  If we concatenate all $D$-dimensional tensor $\mathcal{X}_i$'s  into a $(D+1)$-dimensional tensor $\tilde{\mathcal{X}}=\{ \mathcal{X}_i, i=1,\ldots,N\}=(x_{j_1,\ldots, j_D, i})$, then we consider the CP decomposition of  $\tilde{\mathcal{X}}$ as follows:
 \begin{equation} \label{TPRM2} 
\tilde{\mathcal{X}}= \|\ve\Lambda; {\ve A}^{(1)}, \ldots, {\ve A}^{(D)},{\ve L}\| ~\mbox{or}~ x_{j_1,\ldots,j_D,i}=\sum_{r=1}^{R} \lambda_r a^{(1)}_{j_1r}a^{(2)}_{j_2r} \ldots a^{(D)}_{j_Dr} l_{i r}, 
 \end{equation}  
 where  ${\ve L}=(l_{ir})$ is an $N\times R$ matrix. The matrices ${\ve A}^{(d)}$'s and ${\ve L}$ are the factor matrices.   In this paper,  we introduce the notation  ${\ve L}$  in order to differentiate between matrices that carry common features among subjects (${\ve A}^{(d)}$'s) and the matrix ${\ve L}$,  that is subject specific.

\subsection{ Tensor Partition Regression Models}
\label{model}

\begin{figure} [!htbp]
\begin{center}
\includegraphics[width=0.3\textwidth]{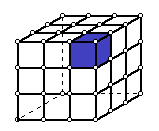}
\caption{ Partition illustration. The purple cube illustrates a sub-tensor $\tilde{\mathcal{X}}^{(s)}$. For $s=1,\ldots,S$, the union of $\tilde{\mathcal{X}}^{(s)}$'s form $\tilde{\mathcal{X}}$, the 3D cube.  }
\label{figurecube}
\end{center}
\end{figure}

  Our interest is to develop TPRM for establishing the association between  responses $y_i$ and  their corresponding imaging covariate $\mathcal{X}_i$ and clinical covariates $\ve z_i$.  
  The first component  of TPRM  is a partition model that divides the   high-dimensional tensor  $\mathcal{X}_i \in \mathbb{R}^{J_1 \times \ldots \times J_D}$  into $S$ disjoint sub-tensor covariates ${\mathcal{X}}_i^{(s)} \in \mathcal{R}^{p_1\times \ldots \times p_D}$ for $s=1, \ldots, S$. Although the size of  ${\mathcal{X}}_i^{(s)}$ can vary across $s$, it is  assumed that,   without loss of generality, 
     ${\mathcal{X}}_i^{(s)}$ and  the  size  of   ${\mathcal{X}}_i^{(s)}$ is homogeneous  such that $S=\prod_{k=1}^D (J_k/p_k)$. We defined the partitions as follows: 
   \begin{gather} \label{TPRM3} 
{\mathcal{X}}_i^{(s)}=\{x_{j_1,j_2,\ldots,j_D i} : j_d \in I_d^{(s_d)}, d=1,\ldots,D\},  \\ 
s=s_1+\sum_{d=2}^D (s_d-1) \{\prod_{k=1}^{d-1}(J_k/p_k)\}, \nonumber \\
1\leq s_d \leq J_d/p_d, ~~
I_d^{(s_d)}=\{(s_d-1)p_d+1, (s_d-1)p_d+2, \ldots, s_dp_d\},  \nonumber\\ \bigcup_{s_d=1}^{J_d/p_d} I_d^{(s_d)}= I_d=\{1,2,\ldots,J_d\} \,\,\, \mbox{and} \,\,\,\,  I_d^{(s_d)} \bigcap I_d^{(s_d')}= \emptyset~\mbox{for}~s_d \neq s'_d. \nonumber  
 \end{gather}   
These sub-tensors ${\mathcal{X}}_i^{(s)}$'s are cubes of neighboring voxels that do not overlap and collectively form the entire 3D image. Figure \ref{figurecube} presents a three-dimensional tensor with sub-tensors.

     The second component of TPRM is a canonical polyadic decomposition model that reduces  the sub-tensor covariates $\tilde{\mathcal{X}}^{(s)}=({\mathcal{X}}_i^{(s)})$ to    low-dimensional feature vectors. Specifically, it is assumed that for each $s$, we have 
 \begin{equation} \label{TPRM4}   
\tilde{\mathcal{X}}^{(s)}=\|\Lambda_{s}; \ve A^{(1)}_{s}, \ve A^{(2)}_{s}, \ldots, \ve A^{(D)}_{s}, \ve L_{s}\| +\mathcal{E}^{(s)}, 
\end{equation} 
where    $\Lambda_{s}=\mbox{diag}(\lambda_1^{(s)}, \ldots, \lambda_R^{(s)})$ consists of the weights for each rank of the decomposition in \eqref{TPRM4},  $\ve A^{(d)}_{s}=(A^{(d)}_{s1} \cdots A^{(d)}_{sr})  \in \mathbb{R}^{p_d\times R}$ is the factor matrix along the $d$-th dimension of $\tilde{\mathcal{X}}^{(s)}$,  and $\ve L_{s}\in \mathbb{R}^{N\times R}$ is the factor matrix along the subject dimension. The error term $\mathcal{E}^{(s)}$ is usually specified in order to find a set of $\ve A^{(d)}_{s}$'s and $\ve L_s$ that best approximates $\tilde{\mathcal{X}}^{(s)}$  \citep{Kolda2009}. We assume that the elements of $\mathcal{E}^{(s)}=(e_{j_1 \ldots j_Di}^{(s)})$ are measurement errors and  $e_{j_1 \ldots j_Di}^{(s)} \sim N(0,(\tau^{(s)})^{-1})$. 
 
 The elements of $\ve L_s$ capture local imaging features in $\mathcal{X}^{(s)}$ across subjects, while  the factor matrix ${\ve A}^{(d)}_s$  represents 
 the common structure of all subjects in  the $d-$th dimension for $d=1,\ldots, D$ 
 \citep{Kolda2009}. In our  ADNI analysis, we have  $D=3$ and ${\ve A}^{(1)}$, ${\ve A}^{(2)}$, and ${\ve A}^{(3)}$ contain the vectors associated with the common features of the images along the coronal, saggital, and axial planes, respectively.
     
     The use of (\ref{TPRM3})  and (\ref{TPRM4}) has two key advantages. First, the partition model (\ref{TPRM3}) allows us to concentrate  on  the most important local features of each sub-tensor, instead of the major variation of the whole image, which may be unassociated with the response of interest. In many applications, although the effect regions (e.g. tumor) associated with responses (e.g. breast cancer) may be relatively small compared with the whole image, their   size   can be comparable with that of each sub-tensor. Therefore, one  can extract more informative features associated with the response with a higher probability.
     Second, the canonical polyadic decomposition model (\ref{TPRM4}) can substantially reduce the dimension of the original imaging data.   For instance, consider a  standard $256\times 256\times 256$ 3D array with 16,777,216 voxels, and its partition model with $32^3=32,768$
 sub-arrays  of size $8\times 8\times 8$. If we reduce each $8\times 8\times 8$ into a small number of components by using  component (ii), then 
 the total number of reduced features is around $O(10^4)$. We can further increase the size of each subarray in order to reduce the size of neuroimaging data  to a manageable level, resulting in efficient estimation.

		The third component of TPRM is a projection of $\ve L=[\ve L_1, \ldots, \ve L_S] \in \mathbb{R}^{N\times P_L}$ ($P_L=S \times R$) into the space spanned by the eigenvectors of $\ve L$. The $i-$th row of $\ve L$, $\ve l_i$ represents the vector of local image features across all partitions. It is assumed that
\begin{equation} \label{TPRM45}
\ve G = \ve L \ve D^T,
\end{equation} where each row of $\ve G$ is a $1\times K$ vector of common unobserved (latent) factors $\ve g_i$ and 
		$\ve D \in \mathbb{R}^{K\times P_L}$ corresponds to the matrix of $K$ basis functions used to represent $\ve L$. Notice that $\ve D$ is  the intrinsic low-dimensional space spanned by all vectors of local image features and,    
   therefore,  $\ve G$ is the projection of $\ve L$ onto $\ve D$. 	
	
	The number of latent basis functions $K$ can be chosen by determining the percentage of data variability in oder to represent $\ve L$ in the basis space. The proposed basis representation has two purposes, including (i) reducing the feature matrix by selecting a small number of basis $K$ and (ii) treating the multicolinearity induced by adjacent partitions in $\ve L$.
     
      The fourth component of TPRM is a generalized linear model that links scalar responses  $y_i$ and  their corresponding reduced  imaging features $\ve g_i$ and clinical covariates $\ve z_i$. Specifically,  
  $ y_i$  given $\ve g_i$ and $\ve z_i$  follows an exponential family distribution with density given by 
 \begin{equation} \label{TPRM5}  
  f(y_i|\ve\theta_i)=m(y_i) \exp\{\eta(\ve\theta_i) T(y_i) -a(\ve \theta_i)\},  
\end{equation} 
  where $m(\cdot)$, $\eta(\cdot)$, $T(\cdot)$, and $a(\cdot)$ are pre-specified functions. Moreover,   it is assumed that 
  $\mu_i = E(y_i|\ve g_i,\ve z_i)$ satisfies 
\begin{equation}    \label{TPRM6}   
  h(\mu_i)= \ve z_i^{T} \ve\gamma + \ve g_i ^{T}\ve b,
\end{equation}
where  $\ve\gamma$ and $\ve b=(b_k)$ are coefficient vectors associated with $\ve z_i$ and $\ve g_i$, respectively,  and $h(\cdot)$ is a link function.

\subsection{Prior Distributions}
\label{priors}

We consider the priors on the elements of $ \ve b$ by assuming a bimodal sparsity promoting prior \citep{ mayrink2013,george1993, george1997} and the following hierarchy: 
\begin{eqnarray}
\label{bprior}
  b_k| \delta_k, \sigma^2 &\sim& (1-\delta_k) F( b_k) + \delta_k \text{N}(0,\sigma^2),  \\
 \delta_k|\pi &\sim& \text{Bernoulli}(\pi)~~~\mbox{and}~~~ 
 \pi \sim \text{Beta}(\alpha_{0\pi},\alpha_{1\pi}), \nonumber
\end{eqnarray}
where $F(\cdot)$ is a pre-specified probability distribution and $\alpha_{0\pi}$ and $\alpha_{1\pi}$ are pre-specified contants. If $F(\cdot)$ is a degenerate distribution at $0$, then  we have the {\itshape spike and slab} prior \citep{Mitchell1988}. A different approach is to consider $F=\text{N}(0,\epsilon)$ with a very small $\epsilon>0$ \citep{Rockova2014}. In this case, the hyperparameter $\sigma^2$ should be large enough to give support to values of  the coefficients that are substantively different from $0$, but not so large that unrealistic values of $ b_k$ are supported. In this article, we opt for the latter approach.

The probability $\pi$ determines whether a particular component of  $\ve g_i$ is informative for predicting $y_i$. A common choice for its hyperparameters is $\alpha_{0\pi}=\alpha_{1\pi}=1$. However, with this choice, the posterior mean of $\pi$ is restricted to the interval $[1/3, 2/3]$, an undesirable feature in variable selection. The `bathtub' shaped beta distribution with $\alpha_{0\pi}=\alpha_{1\pi}=0.5$ concentrates most of its mass in the extremes of the interval $(0,1)$ being more suitable for variable selection \citep{Goncalves2013}.

It is assumed that 
$
 \ve \gamma \sim \mbox{N}(\ve \gamma^*,\upsilon^{-1}\ve I_q)$   and $\upsilon \sim \mbox{Gamma}(\nu_{0\upsilon},\nu_{1\upsilon}),$
 where  $\ve\gamma^*$ is a pre-specified vector and 
  $\nu_{0\upsilon},$ and $\nu_{1\upsilon}$   are pre-specified constants.  
 
If a Bayesian model for the decomposition  \eqref{TPRM4} is selected, we consider the priors on the elements of $A^{(d)}_{sr}$, $\ve l^{(s)}_r$, and $\tau^{(s)}$.    For   $d=1, \ldots, D$ and   $r=1, \ldots, R$,  we assume
\begin{eqnarray} \nonumber 
		A^{(d)}_{sr} \sim \mbox{N}(0,p_d^{-1} {\ve I}_{p_d}),~~~
		\ve l^{(s)}_r \sim \mbox{N}(0, (\tau^{(s)})^{-1}{\ve I}_{N}),~\mbox{and}~
		\tau^{(s)} \sim  \mbox{Gamma}(\nu_{0\tau},\nu_{1\tau}),  
\end{eqnarray}
where $\ve I_N$ is  an $N\times N$ identity matrix and $\nu_{0\tau}$ and $\nu_{1\tau}$ are pre-specified constants. 
 When $p_d$ is large, the columns of the factor matrix $A^{(d)}_{sr}$ are approximately orthogonal, which  is  consistent with their role in the decomposition \eqref{TPRM1} \citep{Ding2011}. However, we do not explicitly require orthonormality, which leads to substantial computational efficiency.

\subsection{Posterior Inference}
\label{poscomp}
Let ${\ve \theta}=\{\ve b,{\ve\delta}, \pi, {\ve\gamma},  \upsilon \} $. A Gibbs sampler algorithm is used to generate a sequence of random observations from the joint posterior distribution given by 
\begin{equation}
\label{posterior}
p(\ve \theta|\mathcal{X},\ve y ) \propto p(\ve y| \ve z, \ve g, \ve\theta)
 p(\ve b|\ve \delta) \,p(\ve\delta|\pi)\, p(\pi) p(\ve\gamma|\upsilon) \,p(\upsilon).  
\end{equation}

The Gibbs sampler essentially involves sampling from a series of conditional distributions,  while each of the modeling components is updated in turn. If the Bayesian model is considered for the tensor decomposition in Equation \eqref{TPRM4}, then ${\ve \theta}=\{ {\ve A}^{(1)},\ldots,{\ve A}^{(D)}, {\ve L},    \ve b,{\ve\delta}, \pi, {\ve\gamma},  \upsilon \} $, where   $\ve\tau=[\tau^{(1)},\ldots, \tau^{(S)}]$. Also, we include  $p(\mathcal{X}|{\ve A}^{(1)}\ldots,{\ve A}^{(D)}, {\ve L}, \ve\tau)$ to the right hand side of \eqref{posterior}.  The detailed sampling algorithm is described in Appendix \ref{app1}.

\section{Simulation Studies}
\label{simsec}

We carried out simulation studies to examine the finite-sample performance of TPRM and its  associated Gibbs sampler. The first study aims at comparing the Bayesian tensor decomposition method with the alternating least squares and to assess the importance of the partition model in the reconstruction of real image. The results are shown in Table \ref{tablermse} of Appendix \ref{appsim}, indicating   that the Bayesian estimation for the tensor components improves the reconstruction error.  However, an important issue associated with using the Bayesian estimation for   \eqref{TPRM4} is its computational burden. For a single  3-dimensional image, running one iteration of the MCMC steps (a.1)$-$(a.4) for a partition of size $33 \times 33 \times 35$ takes 0.72 seconds on a Macintosh OS X, processor 1.4GHz Intel Core i5, memory 8Gb 1600MHz DDR3.   However, when we introduce multiple subjects, as in the examples of the next simulation section and as in the real data application, the computational time  increases to 16 seconds per  iteration even for a single partition. Thus, fitting a full Bayesian TPRM to multiple data sets may become computationally infeasible.  Instead, we calculate the ALS estimates of $\ve L_s$ and then apply MCMC to the fourth component \eqref{TPRM6} of TPRM. This approach is  computationally much more efficient than the full Bayesian  TPRM.

\subsection{A three-dimensional (3D) simulation study}
\label{sec3D}
\begin{figure} [!htbp]
\begin{centering}
\includegraphics[width=12cm]{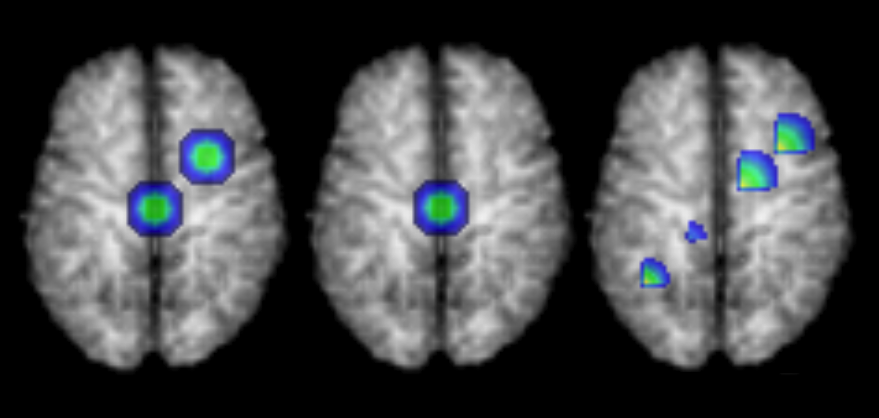}
\caption{The 3D rendering of signal $\mathcal{X}_0$ overlaid on the template $\mathcal{G}_0$ for scenarios 1, 2, and 3, respectively. $\mathcal{X}_0$ is equivalent for scenarios 3 and 4. }
\label{FigSignal3D}
\end{centering}
\end{figure}

The goal of this set of simulations is to examine the classification  performance of the partition model in the 3D imaging setting.  We compare three feature extraction methods including (i) functional principal component model (fPCA); (ii)  tensor alternating least squares (TALS); and (iii) our TPRM. Let  $\mathcal{X}_i\in \mathbb{R}^{64 \times 64\times 50}$ be the image covariate for subject $i$ as defined in Section \ref{notation}. We simulated  $\mathcal{X}_i$'s as follows: 
$$
\mathcal{X}_i(y_i) = \mathcal{G}_0+ y_i\mathcal{X}_0+ \mathcal{E}_i~~~\mbox{for}~~~i=1, \ldots, 200, $$
where $\mathcal{G}_0 \in \mathbb{R}^{64 \times 64\times 50}$ is a fixed brain template with values ranging from $0$ to $250$,
  the elements of the tensor $\mathcal{E}_i \in \mathbb{R}^{64 \times 64\times 50}$ are a noise term, and $\mathcal{X}_0$ is the true  signal image.   Moreover, $\mathcal{X}_0$ is the true  signal image and  was generated according to  the following different scenarios. 
  \begin{enumerate} 
\item[(S.1)] $\mathcal{X}_0$ is composed by two spheres of radius equal to 4 (in voxels)  and the signal decays as it gets farther from their centers; 
\item[(S.2)] $\mathcal{X}_0$ is a sphere of radius equal to 4 (in voxels) and the signal decays as it gets farther from the center of the sphere;
\item[(S.3)] $\mathcal{X}_0=\|{50 \times \ve I}_4; {\ve A^{(1)}},{\ve A^{(2)}}, {\ve A^{(3)}} \|$, where ${\ve A^{(1)}_0} \in \mathbb{R}^{64 \times 4}$, ${\ve A^{(2)}_0} \in \mathbb{R}^{64 \times 4}$, and ${\ve A^{(3)}_0} \in \mathbb{R}^{50\times 4}$, and $\ve A^{(d)}_0$s' are matrices whose $(c_d+j)$-th element of each column is equal to $\sin(j\pi/14)$ with $c_d$ indicating the position at the $d$-th coordinate;
\item[(S.4)] $\mathcal{X}_0 = \|{65 \times \ve I}_4; {\ve A^{(1)}},{\ve A^{(2)}}, {\ve A^{(3)}} \|$, where $\ve A^{(d)}_0$'s  are the same as those in (S.3).  
\item[(S.5)] - (S.8) $\mathcal{X}_0$ is equivalent to scenarios (S.1) - (S.4) except that the elements of $\mathcal{E}_i$ are generated from the short range spacial dependency as described in the first paragraph of this section.
   \end{enumerate} 
   For scenarios (S.1) - (S.4), the elements of $\mathcal{E}_i$ were independently generated from a $\mbox{N}(0,70^2)$ generator.  For scenarios (S.5) - (S.8), the elements of $\mathcal{E}_i=(\mathcal{E}_{i}(g))$ were generated to reflect a short range spatial dependency.  Specifically,  let  $\mathcal{E}_{i}(g) = \sum _{\lVert g'-g \rVert_1 \leq 1} {E}^*_{i}(g')/m_g$, where  $ g$ is a voxel in the three-dimensional space,  
  ${E}^*_{i}(g)\sim \mbox{N}(0,70^2)$,    $\lVert . \rVert_1$ is the $L_1$ norm of a vector, and $m_g$ is the number of locations in the set $\{\lVert g'-g \rVert_1 \leq 1\}$. 
  Figure \ref{FigSignal3D} shows the 3D rendering of $\mathcal{X}_0$  overlaid on the template $\mathcal{G}_0$.

  We consider a specific choice of parameters by setting $R=K=20$ and $S=32$ partitions. Since the signals in $\mathcal{X}_0$ are simple geometric forms,  $20$ basis may be a reasonable choice. In addition, we use the same number of features for all  models to ensure their comparability.  The code for this simulation study is included in the supplemental article \citep{Miranda2017} or can be downloaded from \url{https://github.com/mfmiranda/TPRM}.
 
 With these choices being made, we consider the following criteria. First, we generate the data as described in scenarios (S.1)-(S.8) and split the 200 pairs $( y_i, \mathcal{X}_i)$ into 180 as training samples and 20 as test samples. We perform this splitting 10 times in a 10-fold cross validation procedure. For each combination of training and test set, we use the training set to fit FPCA, TALS, and TPRM. The TPRM model is fitted by  running an MCMC algorithm with 10,000 iterations with a burn-in of 5,000.  The prediction accuracy, the false positive rate and the false negative rate are then computed for each test set. 
 These measurements are the average values across the ten folds for each model  under each scenario. The prediction accuracy (10-fold Accuracy) is the average of the prediction accuracy evaluated at the testing set.
 Results for each scenario and each fold are presented in Tables \ref{Sim32} and \ref{Sim33} of Appendix \ref{simres}. 
 
 Next, we generate 200 pairs $( y_i, \mathcal{X}_i)$, randomly separate them into 180 training samples and 20 test samples, and repeat it 100 times. For each run from $1$ to $100$, we use the training set to fit the models and calculate the prediction accuracy based on the test set. Monte Carlo Accuracy is the average across all these runs.
 
 Table \ref{tablePE} shows the average measurements across multiple runs and also across the ten folds for each model under each scenario. For all scenarios, TPRM  outperforms FPCA and TALS with higher prediction accuracy and smaller FPR and FNR (an exception is the FPR rate for FPCA, since the model is wrongly classifying everyone as positive). For (S.3), the three models are almost equivalent; the prediction accuracy and FNR slightly favor TPRM, but FPR alone favors TALS.

\begin{table}[!htb]
\caption{ A 3D simulation study results for the average prediction accuracy in multiple runs (Monte Carlo Accuracy), follow by the results of a 10-fold cross validation procedure: prediction accuracy (10-fold Accuracy), false positive rate (FPR), and false negative rate (FNR). The partition model TPRM outperforms TALS and FPCA in all scenarios. For Scenario 3, the models are almost equivalent but TPRM is slightly favored. } 
\label{tablePE}
\begin{center}
\begin{tabular}{|c|c|c|c|c|}
\hline
 & &  FPCA &  TALS &   TPRM \\
\hline
 Scenario 1 & Monte Carlo Accuracy  & 0.5615   & 0.5510      &  \textbf{0.8705} \\
  & 10-fold Accuracy  & 0.5750  &  0.5750  &  \textbf{0.8800} \\
 & 10-fold FPR &0  &  0.3750 &   0.1496\\
 & 10-fold FNR & 1.0000 &   0.5081 &   0.0322\\
 \hline
 Scenario 2 & Monte Carlo Accuracy & 0.5795  &  0.5830  &  \textbf {0.8925} \\ 
 & 10-fold Accuracy & 0.5700  &  0.6150  &  \textbf{0.9150} \\
& 10-fold FPR &0.0063 &   0.3817  &  0.0919\\
 & 10-fold FNR & 1.0000  &  0.4494   & 0.0497\\
 \hline
       
 Scenario 3 & Monte Carlo Accuracy &  0.5095  &  0.5330  &  {\bf 0.5710}\\
 & 10-fold Accuracy & 0.5750  &  0.5700  &  \textbf{0.6100} \\
 & 10-fold FPR &0  &  0.2681 &    0.4068\\
 & 10-fold FNR & 1.0000 &   0.6639  &  0.3533\\
 \hline
         
 Scenario 4 & Monte Carlo Accuracy  & 0.5030  &  0.5275  &  {\bf 0.6870} \\
 & 10-fold Accuracy & 0.5750&0.5350 & \textbf{0.7150}  \\
 & 10-fold FPR &0   & 0.3717   & 0.2764\\
 & 10-fold FNR & 1.0000 &   0.5543 &   0.2724\\

 \hline
 Scenario 5 & Monte Carlo Accuracy & 0.7900  &  0.8220     & {\bf 0.9415 } \\
 & 10-fold Accuracy & 0.7600 &   0.8000  & {\bf 0.9250} \\
 & 10-fold FPR &0 &	0.1597 &	0.0918\\
 & 10-fold FNR & 0.5357  &  0.2273  &  0.0245\\
 \hline
 
 Scenario 6 & Monte Carlo Accuracy& 0.6455  &  0.6950 &   {\bf 0.8340}\\
  & 10-fold Accuracy & 0.5850   & 0.7450  &  {\bf 0.8250} \\
 & 10-fold FPR &0 &	0.1457 &   0.1936\\
 & 10-fold FNR & 0.9667   & 0.3730 &   0.1151\\
 
 \hline
 Scenario 7 & Monte Carlo Accuracy& 0.5480  &  0.5480      & {\bf 0.6635} \\
 & 10-fold Accuracy & 0.5750 &   0.5200  & {\bf 0.6750} \\
 & 10-fold FPR &0 &	0.3249 &	0.3427\\
 & 10-fold FNR & 1.0000 &	0.6930 &	0.2651\\
\hline
 Scenario 8 & Monte Carlo Accuracy & 0.6330  &  0.6260     & {\bf 0.7430} \\
  & 10-fold Accuracy & 0.5800 &   0.6550  & {\bf 0.7350} \\
 & 10-fold FPR &0  &  0.2691  &  0.2421\\
 & 10-fold FNR & 0.9833  &  0.4152   & 0.2400\\

 \hline
\end{tabular}  

\end{center}
\end{table}

\section{Real data analysis}
\label{rdsec}

 ``Data used in the preparation of this article were obtained from the Alzheimer's Disease
Neuroimaging Initiative (ADNI) database (adni.loni.usc.edu). The ADNI was launched in
2003 as a public-private partnership, led by Principal Investigator Michael W. Weiner,
MD. The primary goal of ADNI has been to test whether serial magnetic resonance imaging
(MRI), positron emission tomography (PET), other biological markers, and clinical and
neuropsychological assessment can be combined to measure the progression of mild
cognitive impairment (MCI) and early Alzheimer's disease (AD). For up-to-date information, see
{www.adni-info.org}.'' \footnote{ADNI manuscript citation guidelines. 
    {https://adni.loni.usc.edu/wp-content/uploads//how\_to\_apply/ADNI\_DSP\_Policy.pdf}}

We applied the proposed model to the anatomical MRI data collected at the baseline of ADNI. We considered 402 MRI scans from ADNI1, 181 of them were diagnosed with AD ($y_i$ =1), and 221 healthy controls ($y_i$=0). These scans were performed on a 1.5T MRI scanners using a sagittal MPRAGE sequence and the typical protocol includes the following parameters: repetition time (TR) = 2400 ms, inversion time (TI) = 1000 ms, flip angle = 8$^\circ$, and field of view (FOV) = 24 cm with a $256 \times 256 \times 170$ $\mbox{mm}^3$ acquisition matrix in the x, y, and z dimensions, which yields a voxel size of $1.25 \times 1.26 \times 1.2$ $\mbox{mm}^3$ \citep{Huang2014}.

The T1-weighted images were processed using the Hierarchical Attribute Matching Mechanism for Elastic Registration (HAMMER) pipeline. The processing steps include anterior commissure and posterior commissure correction, skull-stripping, cerebellum removal, intensity inhomogeneity correction, and segmentation. Then, registration was performed to
warp the subject to the space of the Jacob template (size $256 \times 256 \times 256$ $\mbox{mm}^3$). Finally, we used the deformation field to compute the RAVENS maps. The RAVENS
methodology precisely quantifies the volume of tissue in each region of the brain. The
process is based on a volume-preserving spatial transformation that ensures that no
volumetric information is lost during the process of spatial normalization \citep{Davatzikos2001}.

\subsection{Functional principal component}

Following the pre-processing steps, we downsampled the images, cropped them, and obtained images of size $96 \times 96 \times 96$ $\mbox{mm}^3$. The simple solution is to consider a classification model, with the response $Y$ being the diagnostics status as described in the previous section, and the design matrix of size $N \times 884,736 (96^3)$. Here, each column of the design matrix is a location in the $3-$D voxel space. Due to the high-dimensionality of the design matrix, we need to consider a dimension reduction approach before fitting a classification model. We consider three classifiers: a classification tree, a support vector machine (SVM) classifier, and a regularized logistic regression with lasso penalty. To evaluate the finite sample performance of the models, we performed a 10-fold cross validation procedure. For each combination of training and test set, we use the training set to extract the first $M$ principal components, with $M$ selected to represent 99\% of the data variability. Next, we use principal components as predictors to fit the models. Then, we evaluate the prediction accuracy on the test set for each data split. The average prediction accuracies across the 10 split sets are $0.6467$, $0.5818$, and $0.5696$ for the tree model, SVM and regularized logistic, respectively. The FPCA approach used here is equivalent to selecting the smallest partition possible (size $1 \times 1\times 1$). In this case, our feature matrix $\ve L$ is formed by all data points in the tensor $\mathcal{X}$. Since the accuracy for these models is low, it is likely that many brain regions associated with the response are not captured by this approach. This limitation highly motivates us to consider the proposed partition model. We believe that finding local features before applying a projection into the principal components space will not only improve prediction accuracy, but also find new and important brain regions that are associated with AD.

\subsection{Selecting the partition model}
We then considered: (i) 64 partitions of size $24\times24\times24$ $\mbox{mm}^3$; (ii) 512 partitions of size $12\times12\times12$ $\mbox{mm}^3$; and (iii) 4096 partitions of size $6\times6\times6$ $\mbox{mm}^3$. For different values of $R$, we selected the number of partitions based on the prediction accuracy of a 10-fold cross validation with the following steps. First, we extracted the features determined by tensor decomposition for different values of rank $R$. Second, to reduce the dimension of the extracted feature matrix, we projected the matrix $\ve L$ into the principal component space with $K$ basis that keeps $90\%$ of the data variability. Third, we run 100,000 iterations of the Bayesian probit model with the mixture prior described in Section \ref{priors} with a burn-in of 5,000 samples, and thinning interval of 50. Finally,  we computed the mean prediction accuracy, the false positive rate and the false negative rate for each data split. Results are shown in Table \ref{tablePA_RD}. We observe that the prediction accuracy does not always increase as $R$ increases. This shows that the locations associated with the response can be represented by a small combination of basis functions. In addition, the accuracy is higher for smaller partitions. This is expected in real data problems when signals  are relatively small and their locations spread throughout the brain.

 \begin{table}[!htb]
 \caption{ ADNI data analysis results: mean prediction accuracy based on 
  a 10-fold cross-validation procedure. The prediction accuracy does not increase as rank increases, and it is bigger for smaller partitions. }
\label{tablePA_RD}
\begin{center}
\begin{tabular}{|c|c|c|c|c|}
\hline
 Partition size & $96\times 96\times 96$ &  $24 \times 24 \times 24 $  &  $12 \times 12 \times 12 $ &  $6 \times 6 \times 6 $ \\
\hline
 $R=5$ &  0.5320 & 0.7040  &     0.6801  &  \textbf{0.9377} \\
 $R=10$& 0.6645 & 0.6670  &     0.8108  &  0.8952 \\
 $R=20$&0.6665 & 0.6791 &     0.8231  &  0.8630 \\
 $R=30$ &0.6492 &0.7418  &     0.8487  &  0.8230 \\
 \hline

\end{tabular}  

\end{center}
\end{table}

\subsection{Final analysis based on the selected model}

\begin{figure} [!htbp]
\includegraphics[width=\textwidth]{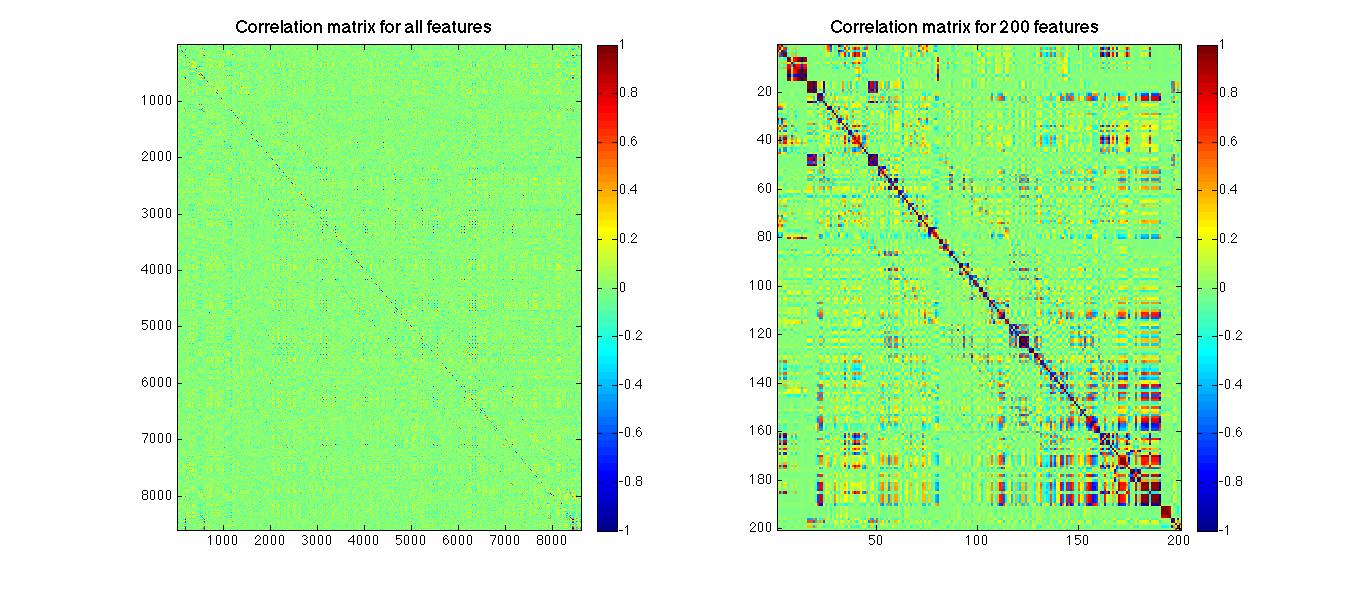}
\caption{ADNI data analysis results. The left panel shows the results for the correlation between the columns of the entire feature matrix  $\ve L$ obtained in the first screening procedure;  the right panel zooms in to shows the same figure for the first 200 features. We observe a high correlation among features in the same partition and among features in neighboring partitions.}
\label{FigCorr}
\end{figure}

Based on the prediction accuracy, we selected the model with all partitions of size $6\times6\times6$ $\mbox{mm}^3$ and $R=5$. For the selected model, we fitted  TPRM with $\sigma^2=10^{4}$, $\epsilon=10^{-4}$, and $\alpha_{0\pi}=\alpha_{1\pi}=0.5$ to reflect the bathtub prior.
In the first screening procedure, we eliminated the partitions whose features, extracted from the tensor decomposition, are zero because they are not relevant in the prediction of  AD. From the 4,096 original partitions, only 1,720 passed the first screening, totaling 8,600 features. Figure \ref{FigCorr} shows the correlation between the features extracted in the first screening step.  Inspecting Figure \ref{FigCorr} reveals    high correlations between features within most partitions and across nearby partitions. Thus, adding the third component of TPRM  can reduce correlation in  the selected features.

Next, we projected the features into the space of principal components. We chose the number of principal components $K$ to enter the final model as follows. Specifically, we  chooe $K$ by specifying the amount of data variation  to be $90\%$.
For this application, we checked the traceplots of the parameters estimates for convergence. The number of final components came down to $K=50$. 

Finally, we run the Gibbs sampler algorithm described in Section \ref{poscomp} for $150,000$ iterations with a burn-in period of $5,000$ iterations and thining interval of 50. Based on a $95\%$ credible interval corrected by the number of test using Bonferroni ($\alpha=0.05/50$), we considered seven components to be important for  predicting  AD outcome. Convergence plots for the 7 coefficients and their correspondent qqplots are shown in Figure \ref{ConvergenceADNI} of Appendix \ref{app2}. Panels (a) to (g) of Figure \ref{FigBasis} present  an axial slice of the 7 important features represented in the image space in their order of importance. The importance is quantified by the absolute value of the posterior mean for each selected feature. We also present a sensitivity analysis for the hyperparameters $\alpha_{0\pi}$ and $\alpha_{1\pi}$ and conclude that the selected features are consistent across different combinations of these hyperparameters. Results are included in Appendix \ref{app0}.

\begin{figure} [!htbp]
\includegraphics[width=\textwidth]{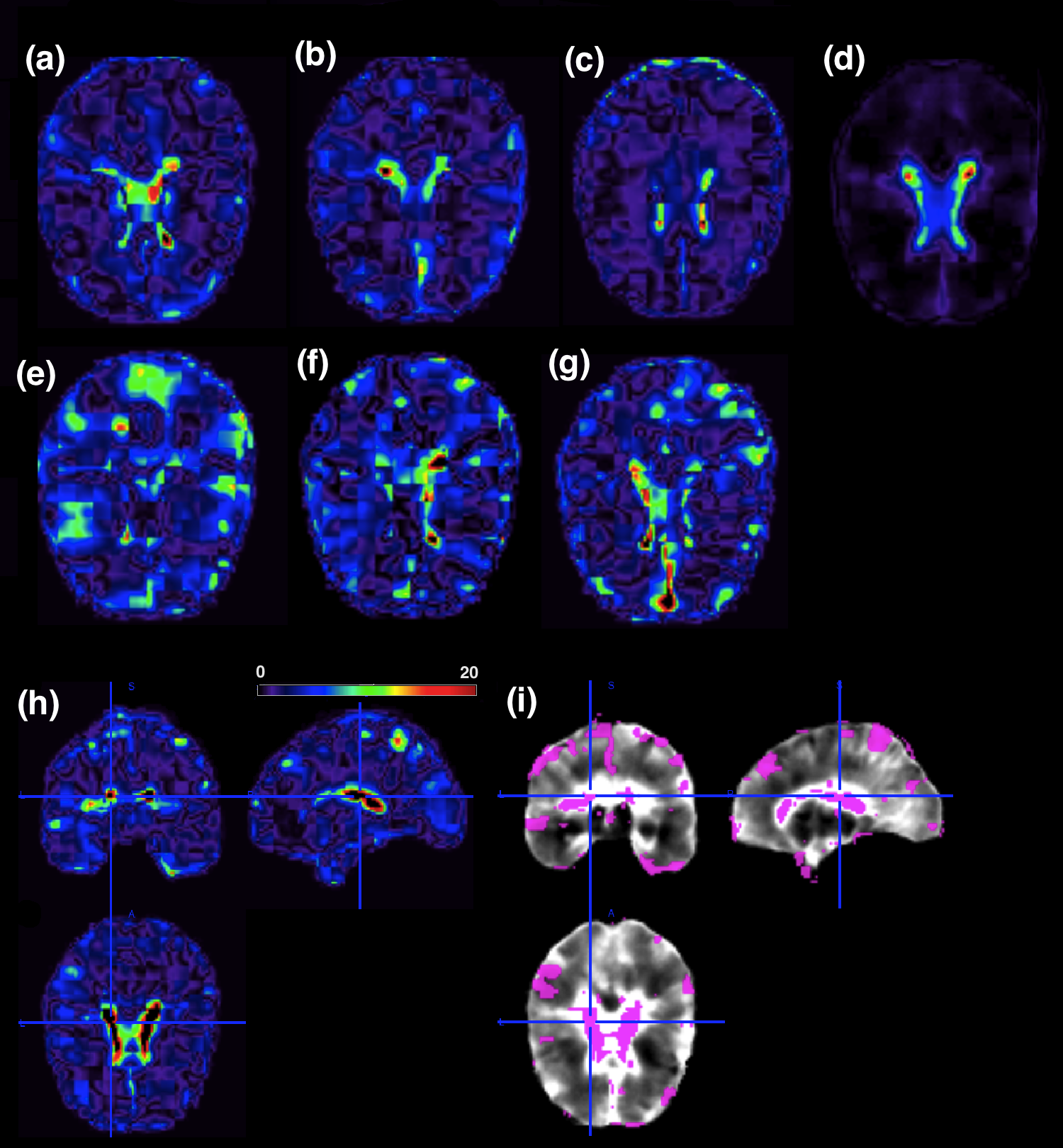}
\caption{ADNI data analysis results: panel (a)-(g) show an axial slice of the most important bases projected into the image space. The importance is given by the absolute value of the posterior mean in each one of the 7 selected features. Panel (h) shows the results for the absolute value of the projection $\mathcal{P}$ for the ADNI dataset.  Colors on the right side of the colorbar indicate regions where differences are higher between the control group and the Alzheimer's group. Panel (i) shows a threshold of $\mathcal{P}$ with colored parts indicating the biomarkers used to predict the onset of AD.}
\label{FigBasis}
\end{figure}

Second, let $\tilde{\ve p}=\hat{\ve b}^{T}{\ve D}$ be a $1 \times P_L$ vector representing the estimated coefficient vector $\hat{\ve b}$ in the local image feature space spanned by the columns of $\ve L$. We computed the projection $\mathcal{P}=  \|\Lambda; {\ve A^{(1)}},{\ve A^{(2)}}, {\ve A^{(3)}},\tilde{\ve p} \|$. The projection $\mathcal{P}$ is a representation of the estimated coefficient vector $\hat{\ve b}$ in the three-dimensional image space. Panel (g) of Figure \ref{FigBasis} presents the absolute value of $\mathcal{P}$, indicating regions of differences  between the control group and the Alzheimer's group. Values on the right hand side of the colorbar are the regions where differences between AD and controls are high. To highlight these biomarkers, we thresholded $\mathcal{P}$ to reveal some of the important regions for AD prediction (Panel (h) of Figure \ref{FigBasis}). The threshold value was chosen to select the $5\%$ highest absolute values of the projection $\mathcal{P}$.

To find specific brain locations that are meaningful for predicting AD, we label the signal locations and present it in Figure \ref{FigBasis} based on  the J\"ulich atlas \citep{Eickhoff2005}. The largest biomarker is the insula, as shown in Table \ref{tableregions}, Appendix \ref{app2}. The insula is associated with perception, self-awareness, and cognitive function.  Many studies have revealed its importance as an AD biomarker \citep{Foundas1997, Karas2004, JackJr2013, Hu2015}. Other important biomarkers are located along the white-matter fiber tracts (fascicles), in particular  a region known as the uncinate fascicle, which contains fiber tracts linking regions of the temporal lobe (such as hippocampus and amygdala) to several frontal cortex regions. Abnormalities within the fiber bundles of the uncinate fasciculus have been previously  associated with AD \citep{Yasmin2008,Salminen2013}.

 Another important biomarker is the hippocampus,  which is associated with learning and consolidation of explicit memories from short-term memory to cortical memory storage for the long term \citep{Campbell2004}. Previous
studies have shown that this region is particularly vulnerable to Alzheimer's disease pathology and already considerably damaged at the time clinical symptoms first appear \citep{Schuff2009, Braak1998}. Other important biomarkers found by TPRM are shown in Table \ref{tableregions},  Appendix \ref{app2}.

\section{Discussion}
\label{dissec}

We have proposed a Bayesian tensor partition regression model (TPRM) to correlate imaging tensor predictors with clinical outcomes. The ultra-high dimensionality of imaging data is dramatically reduced by using the proposed partition model. Our TPRM efficiently addresses the three key features of imaging data, including relatively low signal to noise ratio, spatially clustered effect regions, and the tensor structure of imaging data. 
Our simulations and real data analysis confirm that   TPRM outperforms some state-of-art methods, while  efficiently reducing and identifying relevant imaging biomarkers for accurate prediction.

Many important issues need to be addressed in future research. 
One limitation of TPRM is that the partition tensors are taken from consecutive voxels and therefore do not represent a meaningful brain regions. Such partition is critical for  the tensor decomposition that accounts for  the spatial structure  of medical imaging data. If a prior partition obtained from the existing biological brain regions is preferred, a different basis choice, such as principal components or wavelets, is necessary, since the shapes of these regions will not form a hypercube and therefore tensor decomposition is not applicable.
Another limitation of TPRM is that we only offer an ad hoc approach to select the number of partitions. This approach is not efficient because we have to run many models with different partition sizes in order to identify the best one according to a criterion, such as the prediction accuracy used here. An automated way of selecting the number of partitions is ideal and a topic for future work.

\appendix

\section{Simulation for Bayesian tensor decomposition} 
\label{appsim}

The two goals of the first set of simulations are (i) to compare the Bayesian tensor decomposition method with the alternating least squares method  and (ii) to assess the importance of the partition model in the reconstruction of the original image. 
  We considered 3 different imaging  data sets (or tensors) including  (I$\cdot$1) a diffusion tensor image (DTI) of size $90 \times 96 \times 96$,  (I$\cdot$2) a white matter RAVENS image of size $99 \times 99 \times 70$,   and (I$\cdot$3) a T2-weighted MRI image of size $64 \times 108 \times 99$. We   fitted models (\ref{TPRM3}) and (\ref{TPRM4})  to the three types of  image tensor and  decomposed each of them  with  $R=5,$ $10,$ and $20$. We consider 27 partitions of size $30 \times 30 \times 32$ for the DTI image,    18 partitions of size $33 \times 33 \times 35$ for the RAVENS map, and  24 partitions of size $32 \times 27 \times 33$ for the T2 image, respectively. The hyperparameters $\nu_{0\tau}=1$, $\nu_{1\tau}=10^{-2}$, and $\kappa=10^{-6}$ were chosen to reflect non-informative priors.

\begin{figure} [!htbp]
\includegraphics[width=\textwidth]{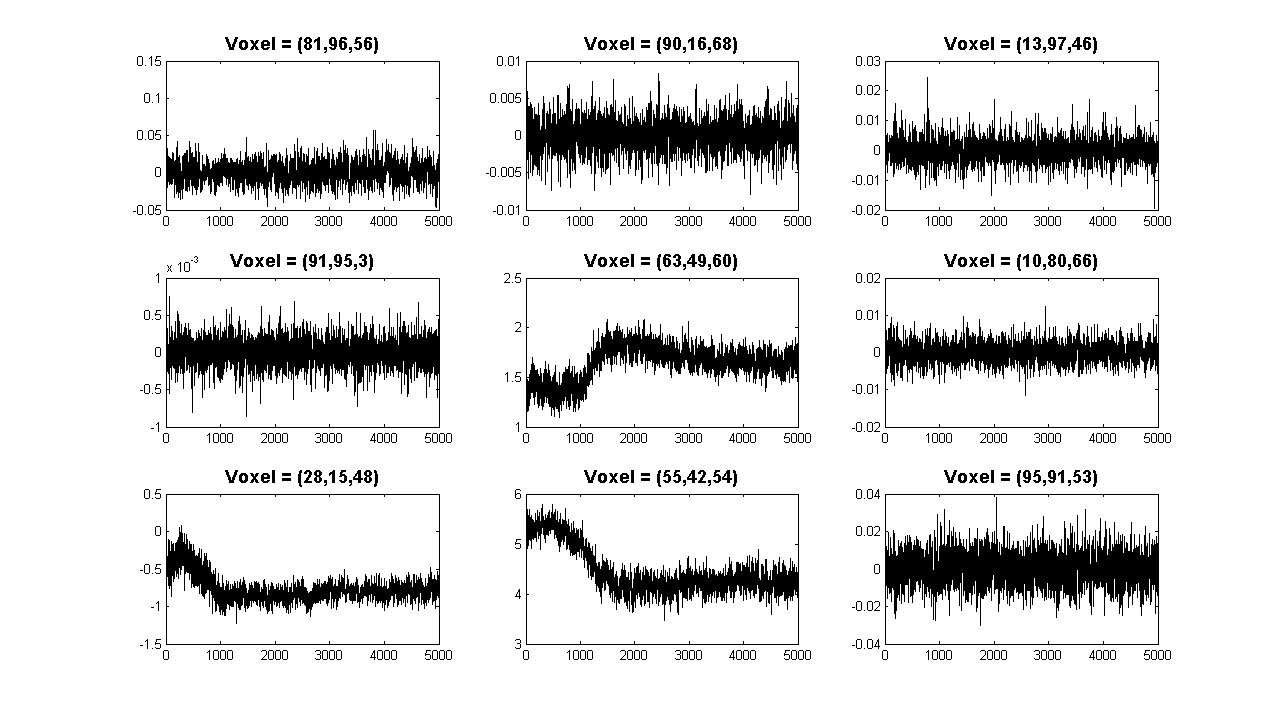}
\caption{Trace plots of Gibbs samplers in 9 randomly selected voxels for  the RAVENS map obtained by Bayesian  tensor decomposition with  $R=20$. The trace plots indicate that the Markov chains converge  after around 2,000 iterations. }
\label{FigDecomp2}
\end{figure}

  We run     steps $(a.1)-(a.4)$ of the Gibbs sampler algorithm in Section \ref{poscomp}   for $5,000$ iterations.  Figure \ref{FigDecomp2} shows the trace plots of Gibbs sampler at  9 randomly selected voxels 
based on       the results for the reconstructed RAVENS map decomposed with $R=20$.  The proposed algorithm   converges very  fast in all voxels. 
  At each iteration,  we computed the quantity $\mathcal{I}=\sum_{s=1}^S \left\|\Lambda_s; \ve A_s^{(1)}, \ve A_s^{(2)}, \ve A_s^{(3)}\right \|$ for each rank and each partition. Subsequently, we computed the reconstructed image, defined as $\mathcal{\hat{X}}$, and the posterior mean estimate of $\mathcal{I}$ after a burn-in sample of $3,000$ iterations. For each reconstructed image $\mathcal{\hat{X}}$, we computed its root mean squared error, $\mbox{RMSE}=||\hat{\mathcal{X}}-\mathcal{X}||_2/\sqrt{J_1J_2J_3}$. 
	
\begin{figure} [!htbp]
\begin{centering}
\includegraphics[width=12cm]{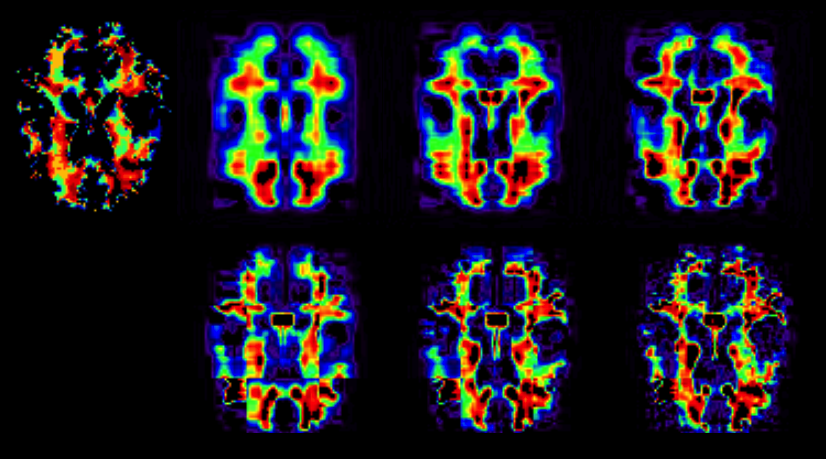}
\caption{Bayesian tensor decomposition results. Top panels: the image on the left represents an axial slice of the RAVENS map image, followed by reconstruction results for the non-partition model. Bottom panels: reconstruction results for the partition model. From left to right,  we have the decomposed images for ranks $R=5,10,$ and $20$, respectively. }
\label{FigDecomp}
\end{centering}
\end{figure}

We consider the non-partition model and compare the Bayesian method with the standard alternating least squares method \citep{Kolda2009}.   
  Figure \ref{FigDecomp} shows an axial slice of  the original white Matter RAVENS map and the reconstructed images for ranks $R=5,10,$ and $20$ as $S=1$.  Table \ref{tablermse} presents RMSEs obtained from the three methods in all scenarios.  The Bayesian decomposition method gives a smaller RMSE for all cases. As expected, the higher the rank, the smaller the reconstruction error.

\begin{table}[!htb]
\caption{Root mean squared error for 3 different types of imaging data. The Bayesian decomposition outperforms the alternating least squares in all scenarios. As the rank $R$ increases, the error decreases. }
\label{tablermse}
\begin{center}
\begin{tabular}{|r|r|c|c|c|}
\hline

           &            & T2-weighted &  WM RAVENS &        DTI \\
\hline

\multicolumn{ 1}{|c|}{R=5} & BayesianCP & 45.3191    &     1.5853 &  3.1656e-004 \\

\multicolumn{ 1}{|c|}{} &        ALS & 45.3636  &     1.6013 &  3.2506e-004 \\

\multicolumn{ 1}{|c|}{} &        Partition & 37.3712  &     1.2178 &   2.0929e-004 \\

\hline
\multicolumn{ 1}{|c|}{R=10} & BayesianCP &  41.7018 &     1.4382 &  2.7367e-004 \\
	
\multicolumn{ 1}{|c|}{} &        ALS &  42.4350    &     1.4533 &  2.8247e-004 \\
\multicolumn{ 1}{|c|}{} &        Partition &  31.3836  &     1.0186 &  1.5748e-004 \\
\hline
\multicolumn{ 1}{|c|}{R=20} & BayesianCP &    37.1796          &   1.2885         &    2.2911e-004         \\

\multicolumn{ 1}{|c|}{} &        ALS &     38.3166       &   1.3166        &    2.3676e-004        \\

\multicolumn{ 1}{|c|}{} &        Partition &    25.1574          &    0.8085         &    1.1349e-004        \\
\hline

\end{tabular}  

\end{center}
\end{table}

\section{Gibbs sampling algorithm for TPRM}\label{app1}

We provide the Gibbs sampling algorithm to sample from the posterior distribution \eqref{posterior} in Section \ref{poscomp}. It involves sampling from a series of conditional distributions,  while each of the modeling components is updated in turn.  
As an illustration,  we divide the whole image into $S$ equal sized regions and assume   $y_i \sim \text{Bernoulli}(\mu_i)$ with the link function $h(\cdot)$ being the probit function.  By following   \cite{Chib1993}, we introduce a normally distributed latent variable, $w_i$, such that 
$
w_i\sim N(\mu_i,  1) $ and $y_i={\bf 1}(w_i>0), 
$
where ${\bf 1}(\cdot)$ is an indicator function of an event.

The complete Gibbs sampler algorithm  proceeds as follows. 
\begin{enumerate}
\item [($a.0$)] Generate $\ve w=(w_1, \ldots, w_n)^T$ from 
\begin{align}
  w_i|y_i=0 &\sim {\bf 1}(w_i\leq 0)\mbox{N}(\ve z_i^{T} \ve\gamma +  \ve g_i^{T}\ve b,1),  \nonumber\\
  w_i|y_i=1 &\sim {\bf 1}(w_i\geq 0)\mbox{N}(\ve z_i^{T} \ve\gamma + \ve g_i^{T}\ve b,1).  \nonumber 
\end{align}

	\item [($a.1$)] Update $\tau(s)$ from its full conditional distribution
	\[ \tau(s)|-  \sim \text{Gamma}(\nu_{0\tau}+(N\prod_{d=1}^D p_d)/2\,,\,\nu_{1\tau}+(1/2)\sum_{i,j_1,\ldots,j_D}(x_{j_1,\ldots,j_Di }^{*}(s))^2),\]
	where $x_{j_1,\ldots,j_Di}^{*}(s)=\{\mathcal{X}^{(s)}-\|\Lambda^{(s)}; {\ve A}^{(1)}_{s}, {\ve A}^{(2)}_{s}, \ldots, {\ve A}^{(D)}_{s}, {\ve L^{(s)}}\|\}_{j_1,\ldots,j_Di}$. 

	\item [($a.2$)] Update $\{{\ve A}^{(d)}_{s}\}_{j_dr}$ from its full conditional distribution given by 
	\[ \{{\ve A}^{(d)}_{s}\}_{j_dr}|-  \sim \text{N}\left(\frac{\tau^{(s)} \langle \widehat{\mathcal{X}}_{(-r)}^{s (j_d)},\mathcal{I}_{(-d)}^s\rangle}{\tau^{(s)}\langle \mathcal{I}_{(-d)}^s,\mathcal{I}_{(-d)}^s\rangle+p_d}, \left(\tau^{(s)}\langle \mathcal{I}_{(-d)}^s,\mathcal{I}_{(-d)}^s\rangle+p_d\right)^{-1}\right), \]
 	where $\mathcal{I}_{(-d)}^s=\|\Lambda^{(s)}; {\ve A}^{(1)}_{s},\ldots, {\ve A}^{(d-1)}_{s}, {\ve A}^{(d+1)}_{s},\ldots,{\ve A}^{(D)}_{s}, {\ve L^{(s)}}\|$,  $\widehat{\mathcal{X}}_{(-r)}^s$ is given by 
$\mathcal{X}^{(s)}-\|\Lambda^{(s)}; {\ve A}^{(1)}_{s}, {\ve A}^{(2)}_{s}, \ldots, {\ve A}^{(D)}_{s}, {\ve L_i^{(s)}}\|+$\\ $\|\Lambda^{(s)}; \{{\ve A}^{(1)}_{s}\}_{:,r}, \{{\ve A}^{(2)}_{s}\}_{:,r}, \ldots, \{{\ve A}^{(D)}_{s}\}_{:,r}, \{{\ve L_i^{(s)}}\}_{:,r}\|, $ and $\widehat{\mathcal{X}}_{(-r)}^{s (j_d)}$ is a subtensor fixed at the entry $j_d$ along the $d$-th dimension of  $\widehat{\mathcal{X}}_{(-r)}^s$.

\item [($a.3$)] Update $\{{\ve L}_{s}\}_{ir}$ from its full conditional distribution given by 
\[	 \{{\ve L}_{s}\}_{ir}|-  \sim \text{N}\left(\frac{\tau^{(s)} \langle \widehat{\mathcal{X}}_{(-r)}^{s (i)},\mathcal{I}^s\rangle}{\tau^{(s)}\langle \mathcal{I}^s,\mathcal{I}^s\rangle+N}, \left(\tau^{(s)}\langle \mathcal{I}^s,\mathcal{I}^s\rangle+N\right)^{-1}\right), \]
 	where $\mathcal{I}^s=\|\Lambda^{(s)}; {\ve A}^{(1)}_{s},\ldots,{\ve A}^{(D)}_{s}\|$, $ \widehat{\mathcal{X}}_{(-r)}^s$ is the same as above,  and $\widehat{\mathcal{X}}_{(-r)}^{s (i)}$ is a subtensor fixed at the $i$-th entry along the subject dimension of  $\widehat{\mathcal{X}}_{(-r)}^s$.
	
	\item[($a.4$)] Normalize the columns of ${\ve A}^{(d)}_{s}$ and $\ve L^{(s)}$ and compute $\Lambda^{(s)}$ with 
	\[ \lambda_r^{(s)}= \|{\ve A}^{(1)}_{s}\| \times \ldots \times \|{\ve A}^{(D)}_{s}\| \times \|{\ve L^{(s)}}\|.\]

\item [($a.5$)] Update $\ve g_k$ from its full conditional distribution
	\begin{eqnarray}\ve g_k|- \sim \mbox{N}(\mu_g,\Sigma_g),~ \Sigma_g=(n\ve I_n+ \tau_{\psi} \sum_{j=1}^{P_L} d_{kj}^2)^{-1} ~
	\mbox{and} ~\mu_g=\tau_{\psi} \Sigma_g \sum_{j=1}^{P_L} d_{kj}{\ve l^{*-k}_j},\nonumber
	\end{eqnarray}
where $\ve l^{*-k}_j={\ve L}-\ve G \ve d_j+ d_{kj}{\ve g_k}$ for $j=1,\ldots, P_L$. 
 
	\item[($a.6$)] Update $ d_{kj}$ 	for $j=1,\ldots, P_L$ from its full conditional distribution
	\[ d_{kj}|- \sim \mbox{N}(\tau_{\psi} \Sigma_d\sum_{j=1}^{P_L} {\ve {g_k^{T}}} {\ve l^{*-k}_j}, \Sigma_d), \]
	where $\Sigma_d=\left(1+ \tau_{\psi} \sum_{j=1}^{P_L} \ve {g_k}^{T} \ve g_k\right)^{-1}$.
	
	\item[($a.7$)] Update $ \tau_{\psi}$ from its full conditional distribution
	\[ \tau_{\psi}|-  \sim \text{Gamma}\left(\beta_{0\psi}+NP_L/2,\beta_{1\psi}+ (\ve L^{*T} \ve L^*)/2\right),\]
		
		where $\ve L^{*}={\ve L}-\ve G \ve D$.
	\item[($a.8$)]  Update $\delta_k$ from its full conditional distribution
	\[\delta_k \sim \mbox{Bernoulli}(\tilde{p}_1/(\tilde{p}_1+\tilde{p}_0)), \]
	where $\tilde{p}_1=\pi \exp\{-(1/2\sigma^2) b_k^2 \}$ and $\tilde{p}_0=\pi \exp\{-(1/2\epsilon)  b_k^2 \}$.
	\item[($a.9$)] Update $\ve b$ from its full conditional distribution

	\begin{align}
	b_k|\delta_k=1 &\sim \mbox{N}(\sum_i \tilde{w}_i g_{ik}/(\sum_i g_{ik}^2+1/\sigma^2),(\sum_i g_{ik}^2+1/\sigma^2)^{-1}),  \nonumber\\
	b_k|\delta_k=0 &\sim \mbox{N}(\sum_i \tilde{w}_i g_{ik}/(\sum_i g_{ik}^2+1/\epsilon),(\sum_i g_{ik}^2+1/\epsilon)^{-1}),  \nonumber
	\end{align}
	 	where $\tilde{w}_i = w_i-\ve z_i^{T} \ve\gamma - \sum_{s'=1}^{S} \ve g_i^{(s')T}\ve b^{(s')}+g_{ir}^{(s)T} b_r^{(s)}$.
	\item[($a.10$)]  Update $\pi$ from its full conditional distribution
	\[ \pi|-  \sim \mbox{beta} (\alpha_{0\pi}+\sum_k \delta_k,\alpha_{1\pi}+K- \sum_{k}\delta_k ). \]
	\item [($a.11$)] Update ${\ve \gamma}$ from its full conditional distribution
	\[ \ve\gamma|-  \sim \text{N}\left(\Sigma_{\gamma}^{*-1} \left(\upsilon \gamma^*+\ve Z^{T} \ve w_{\gamma}^{*}\right),\Sigma_{\gamma}^{*-1}\right),\]
	where $\Sigma_{\gamma}^{*}=\upsilon \ve I_q+\ve Z^T\ve Z$ and $\ve= w_{\gamma}^*= \ve w - \ve G^T\ve b$.
	\item [($a.12$)] Update $\upsilon$ from its full conditional distribution
	\[ \upsilon|-  \sim \text{Gamma}\left(\nu_{0\upsilon}+q/2,\nu_{1\upsilon}+ (\ve\gamma^{T} \ve \gamma)/2\right).\]
\end{enumerate}
All the tensor operations described in steps $(a.1)-(a.4)$ can be easily computed using \cite{TTB_Software}, at {\url{ http://www.sandia.gov/~tgkolda/TensorToolbox/index-2.5.html}.

\newpage

\section{Sensitivity analysis} \label{app0}

We present some results obtained from a sensitivity analysis on the hyperparameters $\alpha_{0\pi}$ and $\alpha_{1\pi}$ in  \eqref{bprior}. For different combinations of the hyperparameters,  we run steps (a.8)-(a.10) in order to  select a  subset of variables. Figure \ref{Sens} shows the MCMC results. The $x$-axis indicates the decision for each of the $K=100$ features. A white color indicates that a specific  feature was selected in TPRM, whereas a black color indicates exclusion. The selected features are similar to each other for all combinations of $\alpha_{0\pi}$ and $\alpha_{1\pi}$.  

\begin{figure} [ht!bp]
\includegraphics[width=\textwidth]{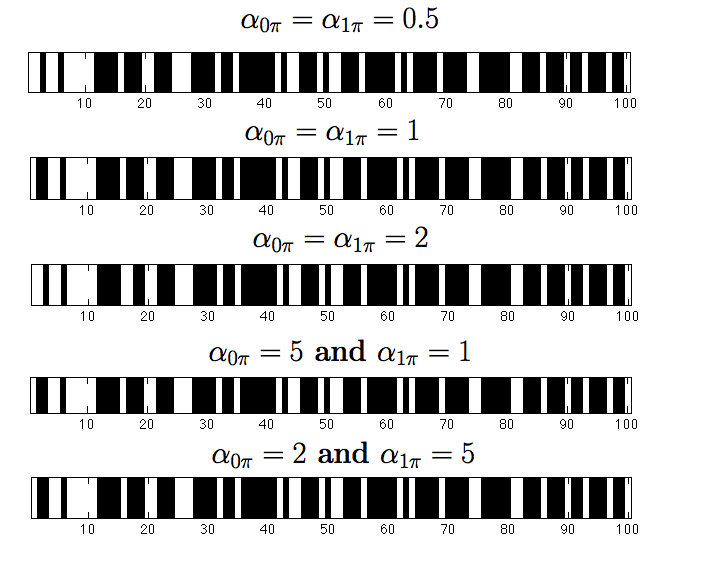}
\caption{Sensitivity analysis for the hyperparameters $\alpha_{0\pi}$ and $\alpha_{1\pi}$ of the bathtub prior in  \eqref{bprior}. A white color indicates that the feature was selected in the model, whereas a black color indicates exclusion. The selected features are similar to each other for all combinations of $\alpha_{0\pi}$ and $\alpha_{1\pi}$.}
\label{Sens}
\end{figure}

\newpage

\section{Real data analysis supporting materials}\label{app2}

\begin{figure} [ht!bp]
\includegraphics[width=1.0\textwidth]{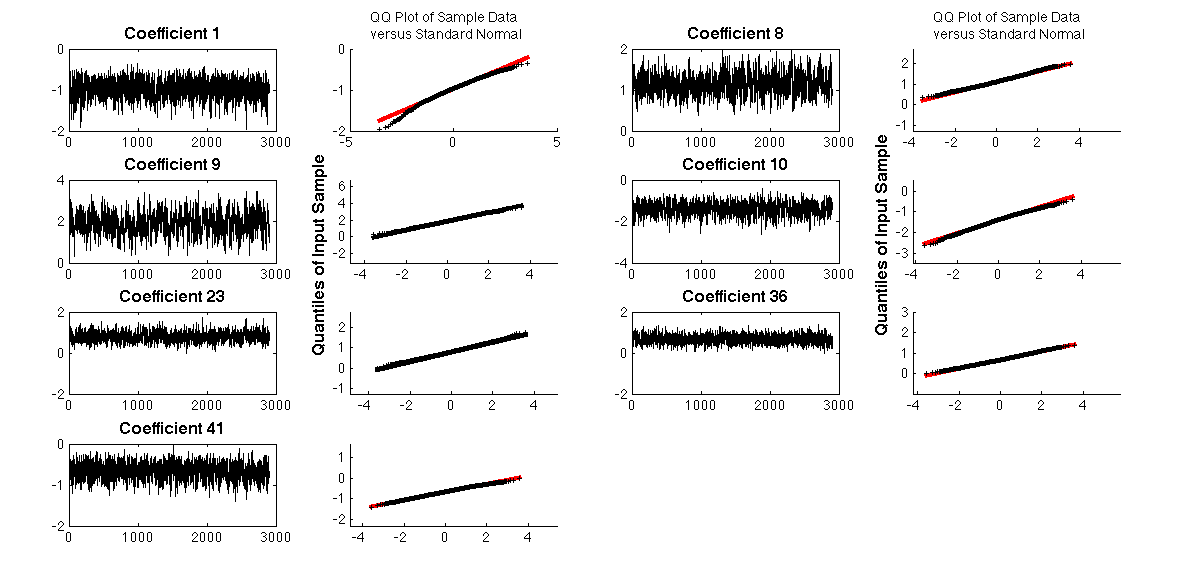}
\caption{Traceplots for the 7 significant coefficients, with their corresponding qqplots. The results confirm  convergence of the MCMC samplers. In addition,  coefficients seem to follow a standard Gaussian distribution.}
\label{ConvergenceADNI}
\end{figure}

\begin{small}

\begin{table} [!htbp]
 \caption{Biomarkers that are relevant to predict AD outcome, based on the J\"ulich atlas. Columns represent the region name, the total amount of voxels in the corresponding region, the number of voxels above the threshold of the projection $\mathcal{P}$, and the percentage of significant voxels considering the total size of the region,  respectively.   }
\label{tableregions}

\begin{tabular}{|l|c|c|c|}
\hline
Region & \# voxels & \# sig. voxels & \% \\
\hline
GM Insula Ig1 R & 189 & 175 & 93 \\
GM Insula Id1 L & 558 & 441 & 79 \\
GM Insula Ig2 R & 743 & 585 & 79 \\
GM Visual cortex V1 BA17 L & 6367 & 4988 & 78 \\
GM Hippocampus dentate gyrus L & 6084 & 4721 & 78 \\
WM Inferior occipito-frontal fascicle L & 1708 & 1305 & 76 \\
GM Superior parietal lobule 7A R & 14507 & 10512 & 72 \\
GM Lateral geniculate body R & 1645 & 1180 & 72 \\
WM Uncinate fascicle L & 571 & 401 & 70 \\
GM Hippocampus dentate gyrus R & 647 & 451 & 70 \\
GM Primary motor cortex BA4a R & 7737 & 5208 & 67 \\
GM Inferior parietal lobule PGp L & 8903 & 5964 & 67 \\
GM Inferior parietal lobule PGp R & 10418 & 6679 & 64 \\
GM Inferior parietal lobule PF R & 7911 & 4957 & 63 \\
GM Broca's area BA44 L & 1555 & 967 & 62 \\
GM Superior parietal lobule 5M R & 2700 & 1668 & 62 \\
GM Primary auditory cortex TE1.0 L & 10423 & 6100 & 59 \\
GM Inferior parietal lobule PFt L & 2054 & 1173 & 57 \\
GM Primary auditory cortex TE1.0 R & 1614 & 895 & 55 \\
GM Primary somatosensory cortex BA1 R & 7170 & 3859 & 54\\
\hline
\end{tabular}

\end{table}
\end{small}

\newpage
\section{Simulation Results, Section 3.1}
\label{simres}

\begin{table}[!htb]
\centering
\begin{scriptsize}
\caption{Model Comparison (FPCA, TALS, and TPRM) - prediction accuracy, false positive rate, and false negative rate for each fold and scenarios (S.1) - (S.4) described on Section \ref{sec3D}.}
\label{Sim32}
\begin{tabular}{|lll|lll|lll|}
\hline
\multicolumn{9}{|c|}{\bf Scenario 1} \\ \hline
\multicolumn{3}{|c|}{Prediction Accuracy} & \multicolumn{3}{c|}{False Positive Rate} & \multicolumn{3}{c|}{False Negative Rate} \\ \hline
 FPCA & TALS & TPRM & FPCA & TALS & TPRM & FPCA & TALS & TPRM \\ \hline 
0.500 & 0.550 & 0.950 & 0.000 & 0.300 & 0.100 & 1.000 & 0.600 & 0.000 \\
0.600 & 0.650 & 0.850 & 0.000 & 0.083 & 0.250 & 1.000 & 0.750 & 0.000 \\
0.350 & 0.450 & 0.850 & 0.000 & 0.714 & 0.000 & 1.000 & 0.462 & 0.231 \\
0.600 & 0.500 & 1.000 & 0.000 & 0.500 & 0.000 & 1.000 & 0.500 & 0.000 \\
0.650 & 0.550 & 0.850 & 0.000 & 0.308 & 0.231 & 1.000 & 0.714 & 0.000 \\
0.800 & 0.650 & 0.700 & 0.000 & 0.438 & 0.375 & 1.000 & 0.000 & 0.000 \\
0.700 & 0.550 & 0.750 & 0.000 & 0.357 & 0.357 & 1.000 & 0.667 & 0.000 \\
0.450 & 0.500 & 0.950 & 0.000 & 0.667 & 0.000 & 1.000 & 0.364 & 0.091 \\
0.500 & 0.650 & 0.950 & 0.000 & 0.300 & 0.100 & 1.000 & 0.400 & 0.000 \\
0.600 & 0.700 & 0.950 & 0.000 & 0.083 & 0.083 & 1.000 & 0.625 & 0.00\\
\hline
\multicolumn{9}{|c|}{\bf Scenario 2} \\ \hline

0.500 & 0.700 & 1.000 & 0.000 & 0.200 & 0.000 & 1.000 & 0.400 & 0.000 \\
0.600 & 0.500 & 0.950 & 0.000 & 0.583 & 0.083 & 1.000 & 0.375 & 0.000 \\
0.350 & 0.700 & 0.900 & 0.000 & 0.571 & 0.000 & 1.000 & 0.154 & 0.154 \\
0.600 & 0.550 & 1.000 & 0.000 & 0.250 & 0.000 & 1.000 & 0.750 & 0.000 \\
0.650 & 0.550 & 0.750 & 0.000 & 0.615 & 0.308 & 1.000 & 0.143 & 0.143 \\
0.750 & 0.600 & 0.800 & 0.063 & 0.375 & 0.250 & 1.000 & 0.500 & 0.000 \\
0.700 & 0.750 & 0.950 & 0.000 & 0.000 & 0.071 & 1.000 & 0.833 & 0.000 \\
0.450 & 0.550 & 1.000 & 0.000 & 0.556 & 0.000 & 1.000 & 0.364 & 0.000 \\
0.500 & 0.700 & 0.800 & 0.000 & 0.500 & 0.200 & 1.000 & 0.100 & 0.200 \\
0.600 & 0.550 & 0.950 & 0.000 & 0.167 & 0.083 & 1.000 & 0.875 & 0.000\\
\hline
\multicolumn{9}{|c|}{\bf Scenario 3} \\ \hline

0.500 & 0.600 & 0.700 & 0.000 & 0.300 & 0.300 & 1.000 & 0.500 & 0.300 \\
0.600 & 0.400 & 0.350 & 0.000 & 0.583 & 0.667 & 1.000 & 0.625 & 0.625 \\
0.350 & 0.300 & 0.850 & 0.000 & 0.714 & 0.000 & 1.000 & 0.692 & 0.231 \\
0.600 & 0.600 & 0.800 & 0.000 & 0.000 & 0.250 & 1.000 & 1.000 & 0.125 \\
0.650 & 0.600 & 0.450 & 0.000 & 0.077 & 0.538 & 1.000 & 1.000 & 0.571 \\
0.800 & 0.650 & 0.700 & 0.000 & 0.375 & 0.313 & 1.000 & 0.250 & 0.250 \\
0.700 & 0.700 & 0.600 & 0.000 & 0.143 & 0.500 & 1.000 & 0.667 & 0.167 \\
0.450 & 0.650 & 0.500 & 0.000 & 0.222 & 0.667 & 1.000 & 0.455 & 0.364 \\
0.500 & 0.600 & 0.550 & 0.000 & 0.100 & 0.500 & 1.000 & 0.700 & 0.400 \\
0.600 & 0.600 & 0.600 & 0.000 & 0.167 & 0.333 & 1.000 & 0.750 & 0.500\\
\hline
\multicolumn{9}{|c|}{\bf Scenario 4} \\ \hline

0.500 & 0.450 & 0.750 & 0.000 & 0.700 & 0.200 & 1.000 & 0.400 & 0.300 \\
0.600 & 0.650 & 0.750 & 0.000 & 0.167 & 0.250 & 1.000 & 0.625 & 0.250 \\
0.350 & 0.450 & 0.750 & 0.000 & 0.143 & 0.143 & 1.000 & 0.769 & 0.308 \\
0.600 & 0.500 & 0.600 & 0.000 & 0.250 & 0.417 & 1.000 & 0.875 & 0.375 \\
0.650 & 0.350 & 0.750 & 0.000 & 1.000 & 0.385 & 1.000 & 0.000 & 0.000 \\
0.800 & 0.600 & 0.650 & 0.000 & 0.438 & 0.375 & 1.000 & 0.250 & 0.250 \\
0.700 & 0.650 & 0.550 & 0.000 & 0.214 & 0.429 & 1.000 & 0.667 & 0.500 \\
0.450 & 0.650 & 0.950 & 0.000 & 0.556 & 0.000 & 1.000 & 0.182 & 0.091 \\
0.500 & 0.550 & 0.600 & 0.000 & 0.000 & 0.400 & 1.000 & 0.900 & 0.400 \\
0.600 & 0.500 & 0.800 & 0.000 & 0.250 & 0.167 & 1.000 & 0.875 & 0.250\\
\hline
\end{tabular}
\end{scriptsize} 
\end{table}

\begin{table}[!htb]
\centering
\begin{scriptsize}
\caption{Model Comparison (FPCA, TALS, and TPRM) - prediction accuracy, false positive rate, and false negative rate for each fold and scenarios (S.5) - (S.8) described on Section \ref{sec3D}.}
\label{Sim33}
\begin{tabular}{|lll|lll|lll|}
\hline
\multicolumn{9}{|c|}{\bf Scenario 5} \\ \hline
\multicolumn{3}{|c|}{Prediction Accuracy} & \multicolumn{3}{c|}{False Positive Rate} & \multicolumn{3}{c|}{False Negative Rate} \\ \hline
 FPCA & TALS & TPRM & FPCA & TALS & TPRM & FPCA & TALS & TPRM \\ \hline 
0.800 & 0.850 & 0.950 & 0.000 & 0.000 & 0.100 & 0.400 & 0.300 & 0.000 \\
0.900 & 0.900 & 1.000 & 0.000 & 0.083 & 0.000 & 0.250 & 0.125 & 0.000 \\
0.350 & 0.750 & 0.900 & 0.000 & 0.143 & 0.000 & 1.000 & 0.308 & 0.154 \\
0.700 & 0.750 & 0.900 & 0.000 & 0.167 & 0.167 & 0.750 & 0.375 & 0.000 \\
0.750 & 0.650 & 0.900 & 0.000 & 0.462 & 0.154 & 0.714 & 0.143 & 0.000 \\
0.900 & 0.800 & 0.850 & 0.000 & 0.188 & 0.188 & 0.500 & 0.250 & 0.000 \\
0.850 & 1.000 & 0.900 & 0.000 & 0.000 & 0.143 & 0.500 & 0.000 & 0.000 \\
0.550 & 0.750 & 0.950 & 0.000 & 0.222 & 0.000 & 0.818 & 0.273 & 0.091 \\
0.850 & 0.750 & 1.000 & 0.000 & 0.000 & 0.000 & 0.300 & 0.500 & 0.000 \\
0.950 & 0.800 & 0.900 & 0.000 & 0.333 & 0.167 & 0.125 & 0.000 & 0.000\\
\hline
\multicolumn{9}{|c|}{\bf Scenario 6} \\ \hline

0.500 & 0.900 & 0.750 & 0.000 & 0.000 & 0.300 & 1.000 & 0.200 & 0.200 \\
0.600 & 0.800 & 0.850 & 0.000 & 0.083 & 0.250 & 1.000 & 0.375 & 0.000 \\
0.350 & 0.900 & 0.750 & 0.000 & 0.000 & 0.143 & 1.000 & 0.154 & 0.308 \\
0.600 & 0.650 & 1.000 & 0.000 & 0.083 & 0.000 & 1.000 & 0.750 & 0.000 \\
0.650 & 0.650 & 0.600 & 0.000 & 0.385 & 0.462 & 1.000 & 0.286 & 0.286 \\
0.800 & 0.850 & 0.750 & 0.000 & 0.188 & 0.313 & 1.000 & 0.000 & 0.000 \\
0.800 & 0.650 & 0.750 & 0.000 & 0.357 & 0.286 & 0.667 & 0.333 & 0.167 \\
0.450 & 0.850 & 0.950 & 0.000 & 0.111 & 0.000 & 1.000 & 0.182 & 0.091 \\
0.500 & 0.650 & 0.900 & 0.000 & 0.000 & 0.100 & 1.000 & 0.700 & 0.100 \\
0.600 & 0.550 & 0.950 & 0.000 & 0.250 & 0.083 & 1.000 & 0.750 & 0.000\\
\hline
\multicolumn{9}{|c|}{\bf Scenario 7} \\ \hline

0.500 & 0.450 & 0.600 & 0.000 & 0.500 & 0.300 & 1.000 & 0.600 & 0.500 \\
0.600 & 0.550 & 0.650 & 0.000 & 0.167 & 0.333 & 1.000 & 0.875 & 0.375 \\
0.350 & 0.350 & 0.650 & 0.000 & 0.429 & 0.286 & 1.000 & 0.769 & 0.385 \\
0.600 & 0.500 & 0.700 & 0.000 & 0.333 & 0.333 & 1.000 & 0.750 & 0.250 \\
0.650 & 0.600 & 0.550 & 0.000 & 0.231 & 0.538 & 1.000 & 0.714 & 0.286 \\
0.800 & 0.650 & 0.750 & 0.000 & 0.375 & 0.313 & 1.000 & 0.250 & 0.000 \\
0.700 & 0.600 & 0.700 & 0.000 & 0.214 & 0.357 & 1.000 & 0.833 & 0.167 \\
0.450 & 0.500 & 0.650 & 0.000 & 0.667 & 0.333 & 1.000 & 0.364 & 0.364 \\
0.500 & 0.550 & 0.750 & 0.000 & 0.000 & 0.300 & 1.000 & 0.900 & 0.200 \\
0.600 & 0.450 & 0.750 & 0.000 & 0.333 & 0.333 & 1.000 & 0.875 & 0.125\\
\hline
\multicolumn{9}{|c|}{\bf Scenario 8} \\ \hline

0.500 & 0.500 & 0.600 & 0.000 & 0.400 & 0.400 & 1.000 & 0.600 & 0.400 \\
0.600 & 0.550 & 0.650 & 0.000 & 0.250 & 0.333 & 1.000 & 0.750 & 0.375 \\
0.350 & 0.750 & 0.600 & 0.000 & 0.143 & 0.286 & 1.000 & 0.308 & 0.462 \\
0.600 & 0.700 & 0.850 & 0.000 & 0.167 & 0.167 & 1.000 & 0.500 & 0.125 \\
0.650 & 0.450 & 0.800 & 0.000 & 0.462 & 0.308 & 1.000 & 0.714 & 0.000 \\
0.800 & 0.500 & 0.700 & 0.000 & 0.625 & 0.375 & 1.000 & 0.000 & 0.000 \\
0.750 & 0.850 & 0.800 & 0.000 & 0.000 & 0.286 & 0.833 & 0.500 & 0.000 \\
0.450 & 0.700 & 0.800 & 0.000 & 0.111 & 0.000 & 1.000 & 0.455 & 0.364 \\
0.500 & 0.800 & 0.800 & 0.000 & 0.200 & 0.100 & 1.000 & 0.200 & 0.300 \\
0.600 & 0.750 & 0.750 & 0.000 & 0.333 & 0.167 & 1.000 & 0.125 & 0.375\\
\hline
\end{tabular}
\end{scriptsize} 
\end{table}

\newpage

\section*{Acknowledgements}
Data collection and sharing for this project was funded by the Alzheimer's Disease
Neuroimaging Initiative (ADNI) (National Institutes of Health Grant U01 AG024904) and
DOD ADNI (Department of Defense award number W81XWH-12-2-0012). ADNI is funded
by the National Institute on Aging, the National Institute of Biomedical Imaging and
Bioengineering, and through generous contributions from the following: Alzheimer’s
Association; Alzheimer’s Drug Discovery Foundation; Araclon Biotech; BioClinica, Inc.;
Biogen Idec Inc.; Bristol-Myers Squibb Company; Eisai Inc.; Elan Pharmaceuticals, Inc.; Eli
Lilly and Company; EuroImmun; F. Hoffmann-La Roche Ltd and its affiliated company
Genentech, Inc.; Fujirebio; GE Healthcare; ; IXICO Ltd.; Janssen Alzheimer Immunotherapy
Research \& Development, LLC.; Johnson \& Johnson Pharmaceutical Research \&
Development LLC.; Medpace, Inc.; Merck \& Co., Inc.; Meso Scale Diagnostics,
LLC.; NeuroRx Research; Neurotrack Technologies; Novartis Pharmaceuticals
Corporation; Pfizer Inc.; Piramal Imaging; Servier; Synarc Inc.; and Takeda Pharmaceutical
Company. The Canadian Institutes of Health Research is providing funds to support ADNI
clinical sites in Canada. Private sector contributions are facilitated by the Foundation for the
National Institutes of Health (www.fnih.org). The grantee organization is the Northern 
Rev December 5, 2013
California Institute for Research and Education, and the study is coordinated by the
Alzheimer's Disease Cooperative Study at the University of California, San Diego. ADNI
data are disseminated by the Laboratory for Neuro Imaging at the University of Southern
California.

\begin{supplement} 
\label{id-suppA}
\stitle{Matlab functions}
\slink[doi]{COMPLETED BY THE TYPESETTER}
\sdatatype{TPRM-Code.zip}
\sdescription{We provide the Matlab code to run the simulation study of Section \ref{sec3D} and the real data in Section \ref{rdsec}}
\end{supplement}
\begin{supplement} 
\stitle{How to obtain the required Matlab toolboxes}
\slink[doi]{COMPLETED BY THE TYPESETTER}
\sdatatype{TPRM-ReadMe.pdf}
\sdescription{We provide the details on how to run the simulation and on how to run TPRM for your own dataset. In addition, we provide information on how to obtain the toolboxes necessary to run the matlab code.}

\end{supplement}

\bibliography{References_AOAS.bib}

\begin{thebibliography}{}

\bibitem[Albert and Chib, 1993]{Chib1993}
Albert, J.~H. and Chib, S. (1993).
\newblock Bayesian analysis of binary and polychotomous response data.
\newblock {\em Journal of the American Statistical Association}, 88(422):pp.
  669--679.

\bibitem[Bader et~al., 2015]{TTB_Software}
Bader, B.~W., Kolda, T.~G., et~al. (2015).
\newblock Matlab tensor toolbox version 2.6.
\newblock Available online.

\bibitem[Bair et~al., 2006]{Bair2006}
Bair, E., Hastie, T., Paul, D., and Tibshirani, R. (2006).
\newblock Prediction by supervised principal components.
\newblock {\em Journal of the American Statistical Association}, 101:119--137.

\bibitem[Beckmann and Smith, 2005]{Beckmann2005}
Beckmann, C.~F. and Smith, S.~M. (2005).
\newblock Tensorial extensions of independent component analysis for
  multisubject f{MRI} analysis.
\newblock {\em NeuroImage}, 25(1):294 -- 311.

\bibitem[Bickel and Levina, 2004]{Bickel2004}
Bickel, P. and Levina, E. (2004).
\newblock Some theory for {F}isher's linear discriminant function, `naive
  {B}ayes', and some alternatives when there are many more variables than
  observations.
\newblock {\em Bernoulli}, 10:989--1010.

\bibitem[Braak and Braak, 1998]{Braak1998}
Braak, H. and Braak, E. (1998).
\newblock Evolution of neuronal changes in the course of {A}lzheimer's disease.
\newblock In Jellinger, K., Fazekas, F., and Windisch, M., editors, {\em Ageing
  and Dementia}, volume~53 of {\em Journal of Neural Transmission.
  Supplementa}, pages 127--140. Springer Vienna.

\bibitem[Breiman et~al., 1984]{breiman84}
Breiman, L., Friedman, J., Olshen, R., and Stone, C. (1984).
\newblock {\em Classification and Regression Trees}.
\newblock Wadsworth, California.

\bibitem[Caffo et~al., 2010]{Caffo2010}
Caffo, B., Crainiceanu, C., Verduzco, G., Joel, S., S.H., M., Bassett, S., and
  Pekar, J. (2010).
\newblock Two-stage decompositions for the analysis of functional connectivity
  for {fMRI} with application to {A}lzheimer's disease risk.
\newblock {\em Neuroimage}, 51(3):1140--1149.

\bibitem[Campbell and MacQueen, 2004]{Campbell2004}
Campbell, S. and MacQueen, G. (2004).
\newblock The role of the hippocampus in the pathophysiology of major
  depression.
\newblock {\em Journal of Psychiatry and Neuroscience}, 29(6):417---426.

\bibitem[Davatzikos et~al., 2001]{Davatzikos2001}
Davatzikos, C., Genc, A., Xu, D., and Resnick, S.~M. (2001).
\newblock Voxel-based morphometry using the {RAVENS} maps: Methods and
  validation using simulated longitudinal atrophy.
\newblock {\em NeuroImage}, 14(6):1361 -- 1369.

\bibitem[Ding et~al., 2011]{Ding2011}
Ding, X., He, L., and Carin, L. (2011).
\newblock Bayesian robust principal component analysis.
\newblock {\em Imaging Processing, IEEE Transactions on}, 20(12):3419--3430.

\bibitem[Eickhoff et~al., 2005]{Eickhoff2005}
Eickhoff, S.~B., Stephan, K.~E., Mohlberg, H., Grefkes, C., Fink, G.~R.,
  Amunts, K., and Zilles, K. (2005).
\newblock A new {SPM} toolbox for combining probabilistic cytoarchitectonic
  maps and functional imaging data.
\newblock {\em NeuroImage}, 25(4):1325 -- 1335.

\bibitem[Fan and Fan, 2008]{FanFan2008a}
Fan, J. and Fan, Y. (2008).
\newblock High-dimensional classification using features annealed independence
  rules.
\newblock {\em Annals of Statistics}, 36:2605--2637.

\bibitem[Foundas et~al., 1997]{Foundas1997}
Foundas, A., Leonard, C., Mahoney, S.~M., Agee, O., and Heilman, K. (1997).
\newblock Atrophy of the hippocampus, parietal cortex, and insula in
  alzheimer's disease: a volumetric magnetic resonance imaging study.
\newblock {\em Neuropsychiatry, Neuropsychology, and Behavioral Neurology},
  10(2):81--9.

\bibitem[Friedman, 1991]{Friedman1991}
Friedman, J. (1991).
\newblock Multivariate adaptive regression splines (with discussion).
\newblock {\em Annals of Statistics}, 19:1--141.

\bibitem[George and McCulloch, 1993]{george1993}
George, E.~I. and McCulloch, R.~E. (1993).
\newblock Variable selection via {G}ibbs sampling.
\newblock {\em Journal of the American Statistical Association}, 88(423):pp.
  881--889.

\bibitem[George and McCulloch, 1997]{george1997}
George, E.~I. and McCulloch, R.~E. (1997).
\newblock Approaches for {B}ayesian variable selection.
\newblock {\em Statistica Sinica}, 7:339--373.

\bibitem[Gillies et~al., 2016]{gillies2015radiomics}
Gillies, R.~J., Kinahan, P.~E., and Hricak, H. (2016).
\newblock Radiomics: Images are more than pictures, they are data.
\newblock {\em Radiology}, 278:563--577.

\bibitem[Gon{\c c}alves et~al., 2013]{Goncalves2013}
Gon{\c c}alves, F., Gamerman, D., and Soares, T. (2013).
\newblock Simultaneous multifactor {DIF} analysis and detection in item
  response theory.
\newblock {\em Computational Statistics \& Data Analysis}, 59(0):144 -- 160.

\bibitem[Hastie et~al., 2009]{Hastie2009}
Hastie, T., Tibshirani, R., and Friedman, J. (2009).
\newblock {\em The Elements of Statistical Learning: Data Mining, Inference,
  and Prediction (2nd)}.
\newblock Springer, Hoboken, New Jersey.

\bibitem[Hu et~al., 2015]{Hu2015}
Hu, X., Meiberth, D., Newport, B., and Jessen, F. (2015).
\newblock Anatomical correlates of the neuropsychiatric symptoms in alzheimer's
  disease.
\newblock {\em Current Alzheimer Research}, 12(3):266--277.

\bibitem[Huang et~al., 2015]{Huang2014}
Huang, M., Nichols, T., Huang, C., Yang, Y., Lu, Z., Feng, Q., Knickmeyere,
  R.~C., Zhu, H., and for~the Alzheimer's Disease Neuroimaging~Initiative
  (2015).
\newblock {FVGWAS}: Fast voxelwise genome wide association analysis of
  large-scale imaging genetic data.
\newblock {\em NeuroImage}, 118:613--627.

\bibitem[Johnstone and Lu, 2009]{Johnstone2009}
Johnstone, I.~M. and Lu, A.~Y. (2009).
\newblock On consistency and sparsity for principal components analysis in high
  dimensions.
\newblock {\em Journal of the American Statistical Assocation}, 104:682--693.

\bibitem[Jr. and Holtzman, 2013]{JackJr2013}
Jr., C. R.~J. and Holtzman, D.~M. (2013).
\newblock Biomarker modeling of alzheimer's disease.
\newblock {\em Neuron}, 80(6):1347 -- 1358.

\bibitem[Karas et~al., 2004]{Karas2004}
Karas, G., Scheltens, P., Rombouts, S., Visser, P., van Schijndel, R., Fox, N.,
  and Barkhof, F. (2004).
\newblock Global and local gray matter loss in mild cognitive impairment and
  alzheimer's disease.
\newblock {\em NeuroImage}, 23(2):708 -- 716.

\bibitem[Kolda, 2006]{kolda2006}
Kolda, T.~G. (2006).
\newblock Multilinear operators for higher-order decompositions.
\newblock Technical report.

\bibitem[Kolda and Bader, 2009]{Kolda2009}
Kolda, T.~G. and Bader, B.~W. (2009).
\newblock Tensor decompositions and applications.
\newblock {\em SIAM Rev.}, 51(3):455--500.

\bibitem[Krishnan et~al., 2011]{Krishnan2011}
Krishnan, A., Williams, L., McIntosh, A., and Abdi, H. (2011).
\newblock Partial least squares ({PLS}) methods for neuroimaging: a tutorial
  and review.
\newblock {\em Neuroimage}, 56:455--475.

\bibitem[Martinez et~al., 2004]{Martinez2004}
Martinez, E., Valdes, P., Miwakeichi, F., Goldman, R.~I., and Cohen, M.~S.
  (2004).
\newblock Concurrent {EEG}/f{MRI} analysis by multiway partial least squares.
\newblock {\em NeuroImage}, 22(3):1023 -- 1034.

\bibitem[Mayrink and Lucas, 2013]{mayrink2013}
Mayrink, V.~D. and Lucas, J.~E. (2013).
\newblock Sparse latent factor models with interactions: Analysis of gene
  expression data.
\newblock {\em The Annals of Applied Statistics}, 7(2):799--822.

\bibitem[Miranda et~al., 2017]{Miranda2017}
Miranda, M.~F., Zhu, H., and Ibrahim, J.~G. (2017).
\newblock Supplement to ``{TPRM}: Tensor partition regression models with
  applications in imaging biomarker detection''.

\bibitem[Mitchell and Beauchamp, 1988]{Mitchell1988}
Mitchell, T.~J. and Beauchamp, J.~J. (1988).
\newblock Bayesian variable selection in linear regression.
\newblock {\em Journal of the American Statistical Association},
  83(404):1023--1032.

\bibitem[M{{\"u}}ller and Yao, 2008]{MR2504202}
M{{\"u}}ller, H.-G. and Yao, F. (2008).
\newblock Functional additive models.
\newblock {\em Journal of the American Statistical Assocation},
  103(484):1534--1544.

\bibitem[Ramsay and Silverman, 2005]{Ramsay2005}
Ramsay, J.~O. and Silverman, B.~W. (2005).
\newblock {\em Functional data analysis}.
\newblock Springer Series in Statistics. Springer, New York, second edition.

\bibitem[Reiss and Ogden, 2010]{Reiss2010}
Reiss, P.~T. and Ogden, R.~T. (2010).
\newblock Functional generalized linear models with images as predictors.
\newblock {\em Biometrics}, 66(1):61--69.

\bibitem[Ro{\v c}kov{\'a} and George, 2014]{Rockova2014}
Ro{\v c}kov{\'a}, V. and George, E.~I. (2014).
\newblock Emvs: The em approach to bayesian variable selection.
\newblock {\em Journal of the American Statistical Association},
  109(506):828--846.

\bibitem[Salminen et~al., 2013]{Salminen2013}
Salminen, L.~E., Schofield, P.~R., Lane, E.~M., Heaps, J.~M., Pierce, K.~D.,
  and Cabeen, R.and~Paul, R.~H. (2013).
\newblock Neuronal fiber bundle lengths in healthy adult carriers of the apoe4
  allele: A quantitative tractography dti study. brain imaging and behavior.
\newblock {\em Brain Imaging and Behavior}, 7(3):81--89.

\bibitem[Schuff et~al., 2009]{Schuff2009}
Schuff, N., Woerner, N., Boreta, L., Kornfield, T., Shaw, L.~M., Trojanowski,
  J.~Q., Thompson, P.~M., Jack~Jr, C.~R., and Weiner, M.~W. (2009).
\newblock {MRI} of hippocampal volume loss in early {A}lzheimer's disease in
  relation to {A}po{E} genotype and biomarkers.
\newblock {\em Brain}, 132(4):1067--1077.

\bibitem[Tibshirani et~al., 2002]{Tibshirani2002}
Tibshirani, R., Hastie, T., Narasimhan, B., and Chu, G. (2002).
\newblock Diagnosis of multiple cancer types by shrunken centroids of gene
  expression.
\newblock {\em Proceedings of the National Academy of Sciences}, 99:6567--6572.

\bibitem[Yasmin et~al., 2008]{Yasmin2008}
Yasmin, H., Nakata, Y., Aoki, S., Abe, O., Sato, N., Nemoto, K., Arima, K.,
  Furuta, N., Uno, M., Hirai, S., Masutani, Y., and Ohtomo, K. (2008).
\newblock Diffusion abnormalities of the uncinate fasciculus in alzheimer's
  disease: diffusion tensor tract-specific analysis using a new method to
  measure the core of the tract.
\newblock {\em Neuroradiology}, 50(4):293--299.

\bibitem[Zhang and Singer, 2010]{Zhang2010}
Zhang, H.~P. and Singer, B.~H. (2010).
\newblock {\em Recursive Partitioning and Applications (2nd)}.
\newblock Springer, New York.

\bibitem[Zhou et~al., 2013]{Zhou2013}
Zhou, H., Li, L., and Zhu, H. (2013).
\newblock Tensor regression with applications in neuroimaging data analysis.
\newblock {\em Journal of the American Statistical Association}, 108:540--552.

\end{thebibliography}

\end{document}